\documentclass[aps,preprint,pre]{revtex4}
\usepackage{epsfig}
\usepackage{graphicx}
\usepackage{color}
\def\bea{\begin{eqnarray}}
\def\eea{\end{eqnarray}}
\def\be{\begin{equation}}
\def\ee{\end{equation}}
\def\mez{\hspace{0.5cm}}
\def\m{\hspace{-1.0mm}}

\def\n{\hspace{-0.25mm}}
\def\be{\begin{equation}}
\def\ee{\end{equation}}
\def\la{\langle}
\def\ra{\rangle}

\newcommand{\dm}{\stackrel{\leftrightarrow}}

\begin{document}
\title{Theory of  diatomic molecules in an external
electromagnetic field from first quantum mechanical principles }

\author{Milan \v{S}indelka and Nimrod Moiseyev}

\affiliation{Department of Chemistry and Minerva Center of
Nonlinear Physics in Complex Systems, Technion -- Israel Institute
of Technology, Haifa 32000, Israel. \\}

\begin{abstract}
We study a general problem of the
translational/rotational/vibrational/electronic dynamics of a
diatomic molecule exposed to an interaction with an arbitrary
external electromagnetic field. The theory developed in this paper
is relevant to a variety of specific applications. Such as,
alignment or orientation of molecules by lasers, trapping of
ultracold molecules in optical traps, molecular optics and
interferometry, rovibrational spectroscopy of molecules in the
presence of intense laser light, or  generation of high order
harmonics from molecules. Starting from the first quantum
mechanical principles, we derive an appropriate molecular
Hamiltonian suitable for description of the center of mass,
rotational, vibrational and electronic molecular motions driven by
the field within the electric dipole approximation. Consequently,
the concept of the Born-Oppenheimer separation between the
electronic and the nuclear degrees of freedom in the presence of
an electromagnetic field is introduced. Special cases of the dc/ac
field limits are then discussed separately. Finally, we consider a
perturbative regime of a weak dc/ac field, and obtain simple
analytic formulas for the associated Born-Oppenheimer
translational/rotational/vibrational molecular Hamiltonian.
\end{abstract}

\maketitle

\section{Introduction}

During the last decade the manipulation of molecules by lasers has
been extensively studied both theoretically and experimentally.
The alignment and orientation of molecules by lasers have a large
variety of applications in different fields of chemistry, physics,
and potentially also in biology and material research
\cite{Articles}. Examples of recently demonstrated applications
range from laser-assisted isotope separation \cite{isotop} and
catalysis \cite{catalysis}, from pulse compression
\cite{Bartels2002} and nanoscale design \cite{Yan1999,Gordon2003}
to tomographic imaging of molecules \cite{Zeidler2004} and quantum
information processing \cite{Lee2004}.

The Hamiltonian for molecules in an external electromagnetic field
is of interest since it underlies a variety of phenomena
associated with the electromagnetic field control of external and
internal molecular motions, including trapping
\cite{Friedrich1995}, molecular optics
\cite{Seideman1997,Yan1999,Sakai1998,Zhao2003,Gordon2003}, Stark
shift manipulation of the potential energy surfaces
\cite{Matusek1996}, and control of the high order harmonic
generation \cite{mol-HG}.

Surprisingly, two qualitatively different forms of the
rovibrational Hamiltonian for molecules in weak laser fields
appear in the theoretical literature dealing with laser alignment.
One form of the Hamiltonian has been derived in
Ref.~\cite{Seideman1999} and the other one in
Ref.~\cite{ARNE_KELLER}. Both approaches, although being
contradictory, have been used extensively in theoretical studies,
giving thus rise to serious confusions and controversies. Most
recently we have resolved this "puzzle" \cite{nimrod-tamar} by
applying the adiabatic theorem for open systems using  an
extension of the $(t,t')$ method, termed the $(t,t',t")$ approach
\cite{Floquet}.

The purpose of this work is to provide a detailed derivation of
the Hamiltonian for diatomic molecules in laser fields, regardless
if the involved field intensities are weak or strong. Our
motivation is to analyze the most general case of a "diatomic
molecule - electromagnetic field" interaction, such that the
obtained results should be relevant not only for description of
molecular alignment, but also for trapping of cold molecules in
optical lattices, for rovibrational spectroscopy of molecules in
the presence of intense laser light, or for generation of high
order harmonics from molecules.

The paper is organized as follows. In Section II, we present a
rigorous self contained derivation of an appropriate molecular
Hamiltonian suitable for description of the center of mass,
rotational, vibrational and electronic molecular motions driven by
the field within the electric dipole approximation. Consequently,
in Section III we introduce the framework of the Born-Oppenheimer
separation between the electronic and the nuclear degrees of
freedom in the presence of an electromagnetic field. Concept of a
time dependent electronic potential energy surface is then
discussed, with particular emphasis on the special cases of the
dc/ac field limits. In Section IV, we consider a perturbative
regime of a weak dc/ac field, and establish an interconnection
between the corresponding Born-Oppenheimer electronic potential
energy surfaces and the conventionally used static/dynamic
molecular polarizabilities. Concluding remarks are given in
Section V.

\section{Diatomic molecule in an electromagnetic field:\\ The
Hamiltonian}

\subsection{The Hamiltonian in momentum gauge and in laboratory frame coordinates }

Let us study a diatomic molecule ${\cal AB}$ exposed to an
interaction with laser light. Some external electrostatic field
can also be present. We prefer here to describe the considered
electromagnetic field classically, in terms of the scalar
potential $\phi(\vec{r},t)$ and the vector potential
$\vec{A}(\vec{r},t)\,$, using Coulomb gauge and Gaussian units for
the electromagnetic quantities \cite{Jackson}. The corresponding
molecular Hamiltonian (expressed with respect to the laboratory
space fixed coordinate frame) possesses an explicit form
\begin{eqnarray} \label{H_sf}
   {\bf H}(t) & = & \frac{1}{2 \, m_{\cal A}} \, \left[ \, \vec{\bf p}_{\cal A} \, - \, \frac{Z_{\cal A} \, e}{c} \,
   \vec{A}(\vec{r}_{\cal A},t) \, \right]^2 \; + \;
   \frac{1}{2 \, m_{\cal B}} \, \left[ \, \vec{\bf p}_{\cal B} \, - \, \frac{Z_{\cal B} \, e}{c} \,
   \vec{A}(\vec{r}_{\cal B},t) \, \right]^2 \; + \; \frac{Z_{\cal A} \, Z_{\cal B} \,
   e^2}{|\vec{r}_{\cal A}-\vec{r}_{\cal B}|} {}\nonumber\\
   & + & \sum_{j=1}^{N} \, \frac{1}{2 \, m_e} \, \left[ \, \vec{\bf p}_j \, + \, \frac{e}{c} \,
   \vec{A}(\vec{r}_j,t) \, \right]^2 \; + \; \sum_{j < j'} \,
   \frac{e^2}{|\vec{r}_j-\vec{r}_{j'}|} {}\nonumber\\
   & - &   \sum_{j=1}^{N} \, \frac{Z_{\cal A} \, e^2}{|\vec{r}_j-\vec{r}_{\cal A}|}
   \; - \; \sum_{j=1}^{N} \, \frac{Z_{\cal B} \, e^2}{|\vec{r}_j-\vec{r}_{\cal B}|}
   \; + \; Z_{\cal A} \, e \, \phi(\vec{r}_{\cal A},t) \; + \; Z_{\cal B} \, e \, \phi(\vec{r}_{\cal B},t)
   \; - \; e \, \sum_{j=1}^{N} \, \phi(\vec{r}_j,t) \mez .
\end{eqnarray}
Here, symbol $e$ stands for a charge of an electron, $c$ denotes
the velocity of light, $Z_{\cal A}$ and $Z_{\cal B}$ are the
atomic numbers of the two nuclei ${\cal A}$ and ${\cal B}$, while
terms $m_{\cal A}, m_{\cal B}$ and $m_e$ represent respectively
the masses of nuclei ${\cal A}, {\cal B}$ or the mass of an
electron. An auxiliary index $j = 1,2,\ldots,N$ has been adopted
for labelling the electronic variables. Other notations should be
self explanatory.

\subsection{The momentum gauge Hamiltonian in the center of mass and relative coordinates}

As the first step of our derivation, we switch from the laboratory
frame coordinates into the center of mass and relative
coordinates. To accomplish this task, we introduce the center of
mass position vector
\be \label{R_c}
   \vec{R}_c \; = \; \frac{ m_{\cal A} \, \vec{r}_{\cal A} \; + \; m_{\cal B} \, \vec{r}_{\cal B}
   \; + \; \sum_{j=1}^{N} \, m_e \, \vec{r}_j}{M} \mez ;
\ee
where the total mass
\be \label{M}
   M \; = \; m_{\cal A} \; + \; m_{\cal B} \; + \; N \, m_e \mez .
\ee
In addition, we define the relative coordinates
\be
\label{R_AB}
   \vec{R}_{{\cal AB}} \; = \; \vec{r}_{\cal A} \; - \; \vec{r}_{\cal B} \mez ;
\ee
and
\be \label{q_j}
   \vec{q}_j \; = \; \vec{r}_j \; - \; \vec{R}_c \mez .
\ee
Relations inverse to the formulas (\ref{R_c}), (\ref{R_AB})
and (\ref{q_j}) are easily found to be
\be \label{r_A}
   \vec{r}_{\cal A} \; = \; \vec{R}_c \; + \; \frac{m_{\cal B}}{m_{{\cal AB}}} \,
   \vec{R}_{{\cal AB}} \; - \; \frac{m_e}{m_{{\cal AB}}} \, \sum_{j=1}^{N} \,
   \vec{q}_j \mez ;
\ee \be \label{r_B}
   \vec{r}_{\cal B} \; = \; \vec{R}_c \; - \; \frac{m_{\cal A}}{m_{{\cal AB}}} \,
   \vec{R}_{{\cal AB}} \; - \; \frac{m_e}{m_{{\cal AB}}} \, \sum_{j=1}^{N} \,
   \vec{q}_j \mez ;
\ee
and \be \label{r_j}
   \vec{r}_j \; = \; \vec{R}_c \; + \; \vec{q}_j \mez .
\ee
An additional auxiliary symbol has been adopted here,
\be
\label{m_AB}
   m_{{\cal AB}} \; = \; m_{\cal A} \; + \; m_{\cal B} \mez .
\ee
For the sake of completeness, we also mention in the present
context that the volume element remains unchanged after the above
described coordinate transformation, i.e.,
\be
\label{volume_element}
   d^3r_{\cal A} \, d^3r_{\cal B} \, \prod_{j=1}^{N} d^3r_j
   \, = \, d^3R_c \; d^3R_{{\cal AB}} \, \prod_{j=1}^{N} d^3q_j \mez .
\ee
Proceeding further, we introduce the momenta associated with
the new coordinates. Namely, we define the operators
\be
\label{momenta}
   \vec{\bf P}_c \; = \; - \, i\hbar \, \nabla_{\vec{R}_c}
   \mez , \mez
   \vec{\bf P}_{{\cal AB}} \; = \; - \, i\hbar \, \nabla_{\vec{R}_{{\cal AB}}}
   \mez , \mez
   \vec{\wp}_j \; = \; - \, i\hbar \, \nabla_{\vec{q}_j} \mez .
\ee
These new momenta are interconnected with the original
laboratory frame momenta through the transformation formulas
\be
\label{p_A}
   \vec{\bf p}_{\cal A} \; = \; \frac{m_{\cal A}}{M} \, \vec{\bf P}_c \; + \;
   \vec{\bf P}_{{\cal AB}} \; - \; \frac{m_{\cal A}}{M} \, \sum_{j=1}^{N} \, \vec{\wp}_j
   \mez ;
\ee \be \label{p_B}
   \vec{\bf p}_{\cal B} \; = \; \frac{m_{\cal B}}{M} \, \vec{\bf P}_c \; - \;
   \vec{\bf P}_{{\cal AB}} \; - \; \frac{m_{\cal B}}{M} \, \sum_{j=1}^{N} \, \vec{\wp}_j
   \mez ;
\ee
and
\be \label{p_j}
   \vec{\bf p}_j \; = \; \frac{m_e}{M} \, \vec{\bf P}_c \; + \; \vec{\wp}_j
   \; - \; \frac{m_e}{M} \, \sum_{j'=1}^{N} \, \vec{\wp}_{j'} \mez .
\ee
We continue by substituting Eqs.~(\ref{R_c}), (\ref{R_AB}),
(\ref{q_j}) and Eqs.~(\ref{p_A}), (\ref{p_B}), (\ref{p_j}) into
Eq.~(\ref{H_sf}). In order to simplify the obtained result, we
employ the dipole approximation
\be \label{dipole_approx}
   \vec{A}(\vec{R}_c+\vec{\xi},t) \; \approx \; \vec{A}(\vec{R}_c,t)
   \mez , \mez \phi(\vec{R}_c+\vec{\xi},t) \; \approx \; \phi(\vec{R}_c,t)
   \; + \; \vec{\xi} \cdot \nabla_{\vec{R}_c}\phi(\vec{R}_c,t) \mez ;
\ee
which is justified as long as the spatial variation of the
electromagnetic field remains negligible at the length scales
$|\vec{\xi}|$ comparable to molecular dimensions.
By using also the transversal property of the Coulomb gauge vector
potential \cite{Jackson}, \be \label{transversality}
   \nabla \cdot \vec{A}(\vec{r},t) \; = \; 0 \mez ;
\ee we write down an explicit expression for the Hamiltonian,
Eq.~(\ref{H_sf}), in the center of mass and relative coordinates.
It holds
\begin{eqnarray} \label{H_sf_cr}
   {\bf H}(t) & = & \frac{\vec{\bf P}_c^2}{2 \, M} \; + \;
   \frac{\vec{\bf P}_{{\cal AB}}^2}{2 \, \mu_{{\cal AB}}} \; + \;
   \sum_{j=1}^{N} \, \frac{\vec{\wp}_j^{\,2}}{2 \, m_e} \; + \;
   \frac{Z_{\cal A} Z_{\cal B} \, e^2}{R_{{\cal AB}}} \; + \; \sum_{j < j'} \,
   \frac{e^2}{|\vec{q}_j-\vec{q}_{j'}|} {}\nonumber\\
   & - & \sum_{j=1}^{N} \, \frac{Z_{\cal A} \, e^2}{\left| \, \vec{q}_j \, - \,
   (m_{\cal B}/m_{{\cal AB}}) \, \vec{R}_{{\cal AB}} \, + \, (m_e/m_{{\cal AB}}) \, \sum_{j'=1}^{N} \vec{q}_{j'}
   \, \right|} {}\nonumber\\
   & - & \sum_{j=1}^{N} \, \frac{Z_{\cal B} \, e^2}{\left| \, \vec{q}_j \, + \,
   (m_{\cal A}/m_{{\cal AB}}) \, \vec{R}_{{\cal AB}} \, + \, (m_e/m_{{\cal AB}}) \, \sum_{j'=1}^{N} \vec{q}_{j'}
   \, \right|} {}\nonumber\\
   & + & \sum_{j=1}^{N} \; \frac{e}{c \, m_e} \, \vec{A}(\vec{R}_c,t) \cdot \vec{\wp}_j
   \; - \; \frac{e}{c} \, \left[ \, \frac{Z_{\cal A}}{m_{\cal A}} \, - \; \frac{Z_{\cal B}}{m_{\cal B}} \, \right]
   \vec{A}(\vec{R}_c,t) \cdot \vec{\bf P}_{{\cal AB}} {}\nonumber\\
   & - & \frac{1}{2 \, M} \sum_{jj'} \, \vec{\wp}_j \cdot \vec{\wp}_{j'} {}\nonumber\\
   & - & \frac{(Z_{\cal A}+Z_{\cal B}-N)}{c \, M} \; \vec{A}(\vec{R}_c,t) \cdot \left[ \,
   \vec{\bf P}_c \, - \, \sum_{j=1}^{N} \, \vec{\wp}_j \, \right] \; + \;
   e \, (Z_{\cal A}+Z_{\cal B}-N) \, \phi(\vec{R}_c,t) {}\nonumber\\
   & + & e \, \Bigl[(Z_{\cal A} m_{\cal B} - Z_{\cal B} m_{\cal A})/m_{{\cal AB}}
   \Bigr] \, \vec{R}_{{\cal AB}} \cdot \nabla_{\vec{R}_c}\phi(\vec{R}_c,t)
   \; - \; e \, \Bigl[ 1 + (Z_{\cal A}+Z_{\cal B})(m_e/m_{{\cal AB}}) \Bigr] \,
   \sum_{j=1}^N \, \vec{q}_j \cdot \nabla_{\vec{R}_c}\phi(\vec{R}_c,t) {}\nonumber\\
   & + & \frac{e^2}{2 \, c^2} \, \left[ \, \frac{Z_{\cal A}^{\,2}}{m_{\cal A}} \, + \, \frac{Z_{\cal B}^{\,2}}{m_{\cal B}}
   \, + \, \frac{N}{m_e} \, \right] \, \vec{A}^{\,2}(\vec{R}_c,t) \mez .
\end{eqnarray}
In Eq.~(\ref{H_sf_cr}) an auxiliary shorthand symbol
\be
\label{reduced_mass}
   \mu_{{\cal AB}} \; = \; \frac{m_{\cal A} \, m_{\cal B}}{m_{\cal A} + m_{\cal B}}
\ee
stands for the reduced mass of the ${\cal AB}$ molecule.
Additional simplifications are in order: {\it i)} The term $(1/(2M))
\sum_{jj'} \, \vec{\wp}_j \cdot \vec{\wp}_{j'}$ in
Eq.~(\ref{H_sf_cr}) can be neglected since the factor $(1/M)$ is
small in magnitude. Similar argument applies also in the case of
terms $(m_e/m_{\cal AB}) \sum_{j=1}^{N} \vec{q}_j\,$. {\it ii)} For
neutral molecules the $[Z_{\cal A} + Z_{\cal B} - N]$-dependent
contributions to Eq.~(\ref{H_sf_cr}) vanish. {\it iii)} The
$\vec{A}^2(\vec{R}_c,t)$ factor can be eliminated from
Eq.~(\ref{H_sf_cr}) by a trivial phase transformation
\cite{length_gauge}, provided that we neglect additional corrections
of the form $M^{-1} [ \, \nabla_{\vec{R}_c} \int^t
\vec{A}^2(\vec{R}_c,t') \, dt' \, ] \cdot \vec{\bf P}_c$ and $M^{-1}
[ \, \Delta_{\vec{R}_c} \int^t \vec{A}^2(\vec{R}_c,t') \, dt' \, ]$
which arise due to non-commutativity between $\vec{A}(\vec{R}_c,t)$
and $\vec{\bf P}_c\,$. This step is justified as long as the spatial
derivatives of the vector potential remain sufficiently small such
that the translational motion of the molecule is not affected by the
mentioned correction terms.

Having incorporated all the above simplifications, we rewrite the
Hamiltonian (\ref{H_sf_cr}) into a relatively simple functional
form
\begin{eqnarray}
\label{H_sf_cr_final}
   {\bf H}(t) & = & \frac{\vec{\bf P}_c^2}{2 \, M} \; + \;
   \frac{\vec{\bf P}_{{\cal AB}}^2}{2 \, \mu_{{\cal AB}}} \; + \;
   \sum_{j=1}^{N} \, \frac{\vec{\wp}_j^{\,2}}{2 \, m_e} \; + \;
   \frac{Z_{\cal A} Z_{\cal B} \, e^2}{R_{{\cal AB}}} \; + \; \sum_{j < j'} \,
   \frac{e^2}{|\vec{q}_j-\vec{q}_{j'}|} {}\nonumber\\
   & - & \sum_{j=1}^{N} \, \frac{Z_{\cal A} \, e^2}{\left| \, \vec{q}_j \, - \,
   (m_{\cal B}/m_{{\cal AB}}) \, \vec{R}_{{\cal AB}} \, \right|} \; - \; \sum_{j=1}^{N} \,
   \frac{Z_{\cal B} \, e^2}{\left| \, \vec{q}_j \, + \, (m_{\cal A}/m_{{\cal AB}}) \, \vec{R}_{{\cal AB}}
   \, \right|} \\
   & + & \sum_{j=1}^{N} \; \frac{e}{c \, m_e} \, \vec{A}(\vec{R}_c,t) \cdot \vec{\wp}_j
   \; - \; \frac{e}{c} \, \left[ \, \frac{Z_{\cal A}}{m_{\cal A}} \, - \; \frac{Z_{\cal B}}{m_{\cal B}} \, \right]
   \vec{A}(\vec{R}_c,t) \cdot \vec{\bf P}_{{\cal AB}} {}\nonumber\\
   & + & e \, \Bigl[(Z_{\cal A} m_{\cal B} - Z_{\cal B} m_{\cal A})/m_{{\cal AB}} \Bigr] \, \vec{R}_{{\cal AB}} \cdot
   \nabla_{\vec{R}_c}\phi(\vec{R}_c,t) \; - \; e \, \sum_{j=1}^N \, \vec{q}_j \cdot
   \nabla_{\vec{R}_c}\phi(\vec{R}_c,t) \mez . {}\nonumber
\end{eqnarray}
The center of mass motion becomes here nonseparable from the
internal molecular motions solely due to presence of the field
terms $\vec{A}(\vec{R}_c,t)$ and $\phi(\vec{R}_c,t)$ in
Eq.~(\ref{H_sf_cr_final}).

\subsection{The length gauge Hamiltonian in the center of mass and relative coordinates}

As the second step of our derivation, we convert the Hamiltonian
(\ref{H_sf_cr_final}) into the length gauge \cite{length_gauge},
which lends itself better for practical applications discussed
later in Section III. The length gauge Hamiltonian $\overline{\bf
H}(t)$ is obtained by an unitary transformation
\be
   \overline{\bf H}(t) \; = \; {\bf U}^\dagger\n(t) \, {\bf H}(t)
   \, {\bf U}(t) \; - \; i\hbar \, {\bf U}^\dagger\n(t) \,
   \frac{\partial}{\partial t} \, {\bf U}(t) \mez ;
\ee
with an unitary operator
\begin{eqnarray}
   {\bf U}(t) & = & \exp\left\{ \, - \, \frac{i}{\hbar} \,
   \frac{e}{c} \; \vec{A}(\vec{R}_c,t) \cdot \sum_{j=1}^{N} \, \vec{q}_j \, + \, \frac{i}{2m} \frac{N e^2}{\hbar c^2} \int^t
   \vec{A}^{\,2}\n(\vec{R}_c,t') \; dt' \, \right\} \\
   & \times & \exp\left\{ \, i \, \frac{e \, \mu_{{\cal AB}}}{\hbar c} \, \left[ \, \frac{Z_{\cal A}}{m_{\cal A}}
   \, - \, \frac{Z_{\cal B}}{m_{\cal B}} \, \right] \, \vec{A}(\vec{R}_c,t) \cdot
   \vec{R}_{{\cal AB}} \, + \, i \, \frac{e^2\mu_{{\cal AB}}}
   {2\hbar c^2} \, \left[ \, \frac{Z_{\cal A}}{m_{\cal A}} \, - \, \frac{Z_{\cal B}}{m_{\cal B}} \, \right]^2
   \int^t \vec{A}^{\,2}\n(\vec{R}_c,t') \; dt' \, \right\} \; . {}\nonumber
\end{eqnarray}
Straightforward algebraic manipulations reveal that
\begin{eqnarray} \label{H_bf_lg}
   \overline{\bf H}(t) & = & \frac{\vec{\bf P}_c^2}{2 \, M} \; + \;
   \frac{\vec{\bf P}_{{\cal AB}}^2}{2 \, \mu_{{\cal AB}}} \; + \;
   \sum_{j=1}^{N} \, \frac{\vec{\wp}_j^{\,2}}{2 \, m_e} \; + \;
   \frac{Z_{\cal A} Z_{\cal B} \, e^2}{R_{{\cal AB}}} \; + \; \sum_{j < j'} \,
   \frac{e^2}{|\vec{q}_j-\vec{q}_{j'}|} {}\nonumber\\
   & - & \sum_{j=1}^{N} \, \frac{Z_{\cal A} \, e^2}{\left| \, \vec{q}_j \, - \,
   (m_{\cal B}/m_{{\cal AB}}) \, \vec{R}_{{\cal AB}} \, \right|} \; - \; \sum_{j=1}^{N} \,
   \frac{Z_{\cal B} \, e^2}{\left| \, \vec{q}_j \, + \, (m_{\cal A}/m_{{\cal AB}}) \, \vec{R}_{{\cal AB}}
   \, \right|} {}\nonumber\\
   & - & \vec{D}_{{\cal AB}}(\vec{R}_{\cal AB},\vec{q}^{\,N}) \cdot \Bigl\{ \, \vec{E}^\parallel\n(\vec{R}_c,t) \, + \,
   \vec{E}^\perp\n(\vec{R}_c,t) \, \Bigr\} \mez .
\end{eqnarray}
Here, the quantity
\be \label{D_AB}
   \vec{D}_{{\cal AB}}(\vec{R}_{\cal AB},\vec{q}^{\,N}) \; = \; e \, \Bigl[(Z_{\cal A} m_{\cal B} - Z_{\cal B} m_{\cal A})/m_{{\cal AB}} \Bigr]
   \, \vec{R}_{{\cal AB}} \; - \; e \, \sum_{j=1}^N \, \vec{q}_j
\ee
can be interpreted as the dipole moment operator of ${\cal
AB}$ molecule, and symbols
\be
   \vec{E}^\perp\n(\vec{R}_c,t) \; = \; - \, \frac{1}{c} \, \frac{\partial
   \vec{A}(\vec{R}_c,t)}{\partial t} \mez , \mez \vec{E}^\parallel\n(\vec{R}_c,t) \; = \;
   - \, \nabla_{\vec{R}_c}\phi(\vec{R}_c,t)
\ee
stand for the transverse and the longitudinal electric fields
assigned to the potentials $\vec{A}(\vec{R}_c,t)$ and
$\phi(\vec{R}_c,t)\,$, respectively \cite{Jackson}. For the sake
of completeness, we note by passing that in the formula
(\ref{H_bf_lg}) we have actually neglected additional corrections
arising due to non-commutativity between the operators
$\vec{A}(\vec{R}_c,t)$ and $\Delta_{\vec{R}_c}$. Justification of
this step is the same as in item {\it iii)} of the previous
subsection II.B.

Before proceeding further in our derivation, let us mention a few
interesting observations regarding the quantity (\ref{D_AB}). For
homonuclear molecules ($m_{\cal A}=m_{\cal B}$ and $Z_{\cal
A}=Z_{\cal B}$) the first term of equation (\ref{D_AB}) vanishes and
thus only the electronic contribution $[-e \sum_{j=1}^N \vec{q}_j]$
is relevant. On the other hand, for cases where $Z_{\cal A}=Z_{\cal
B}=Z$ but $m_{\cal A} \ne m_{\cal B}$ due to the use of different
isotopes (such as HD for example), the formula (\ref{D_AB}) contains
a factor $Z e \, \Bigl[(m_{\cal B} - m_{\cal A})/m_{{\cal AB}}
\Bigr] \, \vec{R}_{{\cal AB}}$ which is acting as a "permanent-like"
dipole moment and influences the photo-induced molecular dynamics.
The mentioned "permanent-like" dipole moment contribution arises in
the case of isotopically substituted homonuclear molecules solely
due to the fact that the nuclear center of mass is not located in
the geometrical center of the ${\cal A} - {\cal B}$ bond (which
constitutes a molecular symmetry center from the point of view of
electronic structure calculations). One might expect that the above
discussed dipole moment component $e \, [(Z_{\cal A} m_{\cal B} -
Z_{\cal B} m_{\cal A})/m_{{\cal AB}} ] \, \vec{R}_{{\cal AB}}$
becomes even more important in the case of heteronuclear diatomics
($m_{\cal A} \neq m_{\cal B}$ and $Z_{\cal A} \neq Z_{\cal B}$).

\subsection{The length gauge Hamiltonian in the spherical polar coordinates}

As the third step of our derivation, we replace the three
cartesian coordinates $\vec{R}_{{\cal AB}} = (X_{{\cal
AB}},Y_{{\cal AB}},Z_{{\cal AB}})$ by their spherical polar
counterparts $(R,\vartheta,\varphi)\,$. We employ the usual
transformation procedure which is well known e.g.~from standard
textbook treatments of the hydrogen atom \cite{Messiah}. The
corresponding transformation formula reads as
\be \label{spc}
   \left( \matrix{ X_{{\cal AB}} \cr Y_{{\cal AB}} \cr Z_{{\cal AB}} } \right) \; = \;
   {\cal M}(\vartheta,\varphi) \, \left( \matrix{ 0 \cr 0 \cr R } \right) \mez ;
\ee
where the rotation matrix
\begin{eqnarray} \label{M_matrix}
   {\cal M}(\vartheta,\varphi) & = &
   \left( \matrix{ +\cos\varphi & -\sin\varphi & 0 \cr
                   +\sin\varphi & +\cos\varphi & 0 \cr
                   0            & 0            & 1    } \right)
   \left( \matrix{ +\cos\vartheta & 0 & +\sin\vartheta \cr
                   0              & 1 & 0              \cr
                   -\sin\vartheta & 0 & +\cos\vartheta    } \right) \; = {}\nonumber\\
   & = &
   \left( \matrix{ \cos\vartheta \cos\varphi & -\sin\varphi & \sin\vartheta \cos\varphi \cr
                   \cos\vartheta \sin\varphi & +\cos\varphi & \sin\vartheta \sin\varphi \cr
                   -\sin\vartheta            & 0            & +\cos\vartheta               }
   \right)
\end{eqnarray}
is orthogonal,
\be \label{M_orthogonal}
   {\cal M} \, {\cal M}^T \; = \; {\cal M}^T \, {\cal M} \; = \;
   {\cal I} \mez .
\ee
The associated volume element is of course $d^3R_{{\cal AB}}
\, = \, R^2 \, dR \, \sin\vartheta \, d\vartheta \, d\varphi\,$.
What remains to be done is to rewrite the Hamiltonian
(\ref{H_bf_lg}) into the new coordinates. An appropriate procedure
for resolving this task is well established, see e.g.~Chapter IX
of Ref.~\cite{Messiah}. Therefore, we display here explicitly just
the final result,
\begin{eqnarray} \label{H_sph}
   \overline{\bf H}(t) & = & \frac{\vec{\bf P}_c^2}{2 \, M} \; + \;
   \frac{{\bf P}_R^2}{2 \, \mu_{{\cal AB}}} \; + \;
   \frac{\vec{\bf L}_{\vartheta\varphi}^2}{2 \, \mu_{{\cal AB}} \, R^2} \; + \;
   \sum_{j=1}^{N} \, \frac{\vec{\wp}_j^{\,2}}{2 \, m_e} \; + \;
   \frac{Z_{\cal A} Z_{\cal B} \, e^2}{R} \; + \; \sum_{j < j'} \,
   \frac{e^2}{|\vec{q}_j-\vec{q}_{j'}|} {}\nonumber\\
   & - & \sum_{j=1}^{N} \, \frac{Z_{\cal A} \, e^2}{\left| \, \vec{q}_j \, - \,
   (m_{\cal B}/m_{{\cal AB}}) \, {\cal M}(\vartheta,\varphi) \, \vec{R}_{{\cal AB}}^{BF} \, \right|} \; - \;
   \sum_{j=1}^{N} \, \frac{Z_{\cal B} \, e^2}{\left| \, \vec{q}_j \, + \, (m_{\cal A}/m_{{\cal AB}}) \,
   {\cal M}(\vartheta,\varphi) \, \vec{R}_{{\cal AB}}^{BF} \, \right|} {}\nonumber\\
   & - & \vec{D}_{{\cal AB}}(R,\vartheta,\varphi,\vec{q}^{\,N}) \cdot \Bigl\{ \, \vec{E}^\parallel\n(\vec{R}_c,t) \, + \,
   \vec{E}^\perp\n(\vec{R}_c,t) \, \Bigr\} \mez .
\end{eqnarray}
Here, the radial momentum operator is defined as
\be \label{P_R}
   {\bf P}_R \; = \; - \, i\hbar \, \frac{1}{R} \, \frac{\partial}{\partial
   R} \, R \mez ;
\ee
the squared angular momentum operator is given by
\be
\label{L2}
   \vec{\bf L}_{\vartheta\varphi}^2 \; = \; - \, \frac{\hbar^2}{\sin^2\vartheta} \, \left[ \,
   \sin\vartheta \, \frac{\partial}{\partial \vartheta} \, \left( \, \sin\vartheta \,
   \frac{\partial}{\partial \vartheta} \, \right) \, + \, \frac{\partial^2}{\partial \varphi^2}
   \, \right] \mez ;
\ee
and an additional auxiliary symbol
\be \label{R_AB_BF}
   \vec{R}_{{\cal AB}}^{BF} \; = \; [ \, 0 , 0 , R \, ] \mez .
\ee
To avoid confusion, let us note explicitly that the dipole
moment operator (\ref{D_AB}) is now expressed in the form
\be
\label{D_AB_sph}
   \vec{D}_{{\cal AB}}(R,\vartheta,\varphi,\vec{q}^{\,N}) \; = \; e \, \Bigl[(Z_{\cal A} m_{\cal B} - Z_{\cal B} m_{\cal A})/m_{{\cal AB}} \Bigr]
   \, {\cal M}(\vartheta,\varphi) \, \vec{R}_{{\cal AB}}^{BF} \; - \; e \, \sum_{j=1}^N \, \vec{q}_j \mez .
\ee

\subsection{The length gauge Hamiltonian in the body
fixed electronic coordinates}

In this step of our derivation, we transform the position vectors
of all the electrons into the body fixed frame. The
origin $O$ of the body fixed coordinate system is set to be the
molecular center of mass. Note that this choice of the origin 
is a bit different from the choice adopted within the usual 
spectroscopic literature, where the nuclear center of mass is considered 
instead (see for example Ref.~\cite{Carrington}). We prefer to use here
an alternative less conventional assignment of the origin $O$ since it 
makes our formulation more transparent and enables us to avoid introducing additional 
approximations.

The body fixed $\oplus \, o_z$ axis is,
by definition, parallel (although not always coincidental) with the direction of $\vec{R}_{{\cal
AB}}\,$. The body fixed $o_x$ and $o_y$ axes are constrained by
the requirement $\oplus \, o_x \; \times \; \oplus \, o_y \; = \;
\oplus \, o_z\,$. Choice of $o_x$ and $o_y$ is, however, not
unique: An arbitrary rotation around $o_z$ leads to an equivalent
pair of body fixed axes $(o'_x,o'_y)$ which are equally suitable
as $(o_x,o_y)\,$. Hence, an unambiguous definition of $o_x$ and
$o_y$ must be fixed by convention. In order to achieve maximum
simplicity, we prefer to employ such a particular convention that
\be \label{r_j_bf}
   \vec{q}_j \; = \; {\cal M}(\vartheta,\varphi) \; \vec{\tt r}_j
   \mez ;
\ee
where $\vec{\tt r}_j$ are the body fixed coordinates of vector
$\vec{q}_j\,$.  Since the matrix ${\cal M}(\vartheta,\varphi)$ is
orthogonal, the volume element remains unaffected, $d^3q_j \, = \,
d^3{\tt r}_j\,$. Having introduced the body fixed electronic
coordinates, we continue further and define the associated
momenta,
\be \label{p_j_bf}
   \vec{\tt p}_j \; = \; - \, i\hbar \, \nabla_{\vec{\tt r}_j}
   \mez .
\ee
These new momenta are interconnected with their space fixed
counterparts through the transformation formulas
\be
\label{p_j_sf_bf}
   \vec{\wp}_j \; = \; {\cal M}(\vartheta,\varphi) \; \vec{\tt p}_j
   \mez .
\ee
It is straightforward to rewrite the Hamiltonian (\ref{H_sph})
into the body fixed coordinates. Taking advantage of the
orthogonality property (\ref{M_orthogonal}), we arrive towards the
desired result
\begin{eqnarray} \label{H_bf}
   \overline{\bf H}(t) & = & \frac{\vec{\bf P}_c^2}{2 \, M} \; + \;
   \frac{{\bf P}_R^2}{2 \, \mu_{{\cal AB}}} \; + \;
   \frac{\vec{\bf L}_{\vartheta\varphi}^2}{2 \, \mu_{{\cal AB}} \, R^2} \; + \;
   \sum_{j=1}^{N} \, \frac{\vec{\tt p}_j^{\,2}}{2 \, m_e} \; + \;
   \frac{Z_{\cal A} Z_{\cal B} \, e^2}{R} \; + \; \sum_{j < j'} \,
   \frac{e^2}{|\vec{\tt r}_j-\vec{\tt r}_{j'}|} {}\nonumber\\
   & - & \sum_{j=1}^{N} \, \frac{Z_{\cal A} \, e^2}{\sqrt{{\tt x}_j^2+{\tt y}_j^2+
   [\,{\tt z}_j-(m_{\cal B}/m_{{\cal AB}})R\,]^2}} \; - \;
   \sum_{j=1}^{N} \, \frac{Z_{\cal B} \, e^2}{\sqrt{{\tt x}_j^2+{\tt y}_j^2+
   [\,{\tt z}_j+(m_{\cal A}/m_{{\cal AB}})R\,]^2}} {}\nonumber\\
   & - & \Bigl[ \, {\cal M}(\vartheta,\varphi) \, \vec{D}_{{\cal AB}}^{BF}(R,\vartheta,\varphi,\vec{\tt r}^{\,N}) \, \Bigr] \cdot
   \Bigl\{ \, \vec{E}^\parallel\n(\vec{R}_c,t) \, + \, \vec{E}^\perp\n(\vec{R}_c,t) \, \Bigr\}
   \mez .
\end{eqnarray}
Here, the quantity
\be \label{D_BF}
   \vec{D}_{{\cal AB}}^{BF}(R,\vartheta,\varphi,\vec{\tt r}^{\,N}) \; = \; e \, \Bigl[(Z_{\cal A} m_{\cal B} - Z_{\cal B} m_{\cal A})/m_{{\cal AB}} \Bigr]
   \, \vec{R}_{{\cal AB}}^{BF} \; - \; e \, \sum_{j=1}^N \, \vec{\tt r}_j
\ee
represents the body fixed counterpart of the dipole moment
vector (\ref{D_AB}). Note that in formula (\ref{H_bf}) the
translational and rovibrational kinetic energy operators are
completely decoupled from the kinetic energy operators of the
electrons. This holds true in particular also for the electronic and
the nuclear angular momenta. (The electronic angular momenta are not
displayed here explicitly and appear only after switching into the
spherical or cylindrical electronic coordinates.)


\subsection{Final form of the Hamiltonian for diatomic molecules in an electromagnetic field}

Summarizing all the elaborations of Section II, we may conclude
that the quantum dynamics of the considered molecule ${\cal AB}$
interacting with an external electromagnetic field
$\vec{A}(\vec{r},t)$ and $\phi(\vec{r},t)$ is described by the
time dependent Schr\"{o}dinger equation
\be \label{SCHE_1}
   i\hbar \, \frac{\partial}{\partial t} \, \Xi(\vec{R}_c,R,\vartheta,\varphi,\vec{\tt r}^N\m,t)
   \; = \; \overline{\bf H}(t) \; \Xi(\vec{R}_c,R,\vartheta,\varphi,\vec{\tt
   r}^N\m,t) \mez ;
\ee
where $\Xi(\vec{R}_c,R,\vartheta,\varphi,\vec{\tt r}^N\m,t)$ is
the associated translational/rotational/vibrational/electronic
wavefunction, and the appropriate Hamiltonian $\overline{\bf H}(t)$
is given by expression (\ref{H_bf}). To avoid confusion, let us note
in passing that the electron spin variables are suppressed in the
notation of the present paper, since they never enter explicitly
into our considerations. Nevertheless, the presence of an electronic
spin is of course fully respected within our treatment, as well as
the antisymmetry of the electronic wavefunctions.

Before proceeding further, it is convenient to introduce an
additional simplification, based upon the factorization \be
\label{factorization}
   \Xi(\vec{R}_c,R,\vartheta,\varphi,\vec{\tt r}^N\m,t) \; = \;
   R^{-1} \, \Psi(\vec{R}_c,R,\vartheta,\varphi,\vec{\tt r}^N\m,t)
   \mez .
\ee
The purpose of this factorization is to eliminate redundant
difficulties arising due to a complicated functional form of the
radial momentum (\ref{P_R}). We refer again to standard textbooks
\cite{Messiah} for a more detailed discussion of this issue. One
can easily show that the redefined wavefunction
$\Psi(\vec{R}_c,R,\vartheta,\varphi,\vec{\tt r}^N\m,t)$ satisfies
the time dependent Schr\"{o}dinger equation
\be \label{SCHE_2}
   i\hbar \, \frac{\partial}{\partial t} \, \Psi(\vec{R}_c,R,\vartheta,\varphi,\vec{\tt r}^N\m,t)
   \; = \; \tilde{\bf H}(t) \, \Psi(\vec{R}_c,R,\vartheta,\varphi,\vec{\tt r}^N\m,t) \mez ;
\ee
with the Hamiltonian
\be \label{H_full_simple}
   \tilde{\bf H}(t) \; = \; - \, \frac{\hbar^2}{2 M} \, \Delta_{\vec{R}_c} \; - \;
   \frac{\hbar^2}{2 \mu_{{\cal AB}}} \, \frac{\partial^2}{\partial R^2} \; + \;
   \frac{\vec{\bf L}_{\vartheta\varphi}^2}{2 \, \mu_{{\cal AB}} \, R^2} \; + \;
   {\bf H}_{\rm el}(R) \; + \; {\bf W}_{\rm el}(\vec{R}_c,R,\vartheta,\varphi,t) \mez .
\ee
Here, the field free electronic Hamiltonian
\begin{eqnarray}
\label{H_el_FF}
   \hspace*{-0.50cm} {\bf H}_{\rm el}(R) & = &
   \sum_{j=1}^{N} \, \frac{\vec{\tt p}_j^{\,2}}{2 \, m_e} \; + \;
   \frac{Z_{\cal A} Z_{\cal B} \, e^2}{R} \; + \; \sum_{j < j'} \,
   \frac{e^2}{|\vec{\tt r}_j-\vec{\tt r}_{j'}|} {}\nonumber\\
   \hspace*{-0.50cm} & - & \sum_{j=1}^{N} \, \frac{Z_{\cal A} \, e^2}{\sqrt{{\tt x}_j^2+{\tt y}_j^2+
   [\,{\tt z}_j-(m_{\cal B}/m_{{\cal AB}})R\,]^2}} \; - \;
   \sum_{j=1}^{N} \, \frac{Z_{\cal B} \, e^2}{\sqrt{{\tt x}_j^2+{\tt y}_j^2+
   [\,{\tt z}_j+(m_{\cal A}/m_{{\cal AB}})R\,]^2}} \mez ;
\end{eqnarray}
the "${\cal AB}$ molecule - field" interaction term \be
\label{W_el}
   {\bf W}_{\rm el}(\vec{R}_c,R,\vartheta,\varphi,t) \; = \; - \, \Bigl[ \,
   {\cal M}(\vartheta,\varphi) \, \vec{D}_{{\cal AB}}^{BF}(R,\vartheta,\varphi,\vec{\tt r}^{\,N}) \, \Bigr] \cdot \vec{E}(\vec{R}_c,t) \mez ;
\ee and an overall electric field \be
   \vec{E}(\vec{R}_c,t) \; = \; \vec{E}^\parallel\n(\vec{R}_c,t) \; + \;
   \vec{E}^\perp\n(\vec{R}_c,t) \mez .
\ee


\section {Diatomic molecule in an
electromagnetic field:\\ The time-dependent Born-Oppenheimer
electronic potential\\ energy surfaces}

\subsection{The time-dependent electronic wavefunctions}

For isolated molecules, the well known concept of the
Born-Oppenheimer/adiabatic separation between the electronic and the
nuclear degrees of freedom proved to be extremely useful, as it
gives a lot of physical insight by distinguishing between the
electronic and the rovibrational molecular states. The purpose of
this Section is to extend the formulation of the Born-Oppenheimer
approach on cases when a molecule is exposed to an interaction with
an electromagnetic field. For simplicity, we shall consider here
just a diatomic molecule ${\cal AB}$ for which an appropriate
Hamiltonian has been discussed at length in the previous Section II.

As the first step of our analysis, we formally construct the time
dependent Born-Oppenheimer electronic basis set, defined by
particular solutions of the electronic time dependent
Schr\"{o}dinger equation
\be \label{SCHE_el}
   i\hbar \, \frac{\partial}{\partial t} \,
   \Phi_n(\vec{\tt r}^N\m,t;\vec{R}_c,R,\vartheta,\varphi)
   \; = \; \Bigl[ \, {\bf H}_{\rm el}(R) \, + \,
   {\bf W}_{\rm el}(\vec{R}_c,R,\vartheta,\varphi,t) \, \Bigr] \,
   \Phi_n(\vec{\tt r}^N\m,t;\vec{R}_c,R,\vartheta,\varphi) \mez .
\ee
The corresponding initial condition is conveniently chosen to
be
\be \label{intcond}
   \Phi_n(\vec{\tt r}^N\m,t_0;\vec{R}_c,R,\vartheta,\varphi) \; =
   \; \Phi_n^0(\vec{\tt r}^N\m;R) \mez ;
\ee
where $t_0$ is an as yet arbitrary time instant, and function
$\Phi_n^0(\vec{\tt r}^N\m;R)$ represents a solution of a field
free electronic eigenproblem
\be \label{evf_el_FF}
   {\bf H}_{\rm el}(R) \, \Phi_n^0(\vec{\tt r}^N\m;R) \; = \;
   {\cal E}_n^0(R) \, \Phi_n^0(\vec{\tt r}^N\m;R) \mez .
\ee
In order to simplify the underlying notation, we assume here
that the electronic Hamiltonian ${\bf H}_{\rm el}(R)$ possesses a
discrete spectrum labelled by a single collective index $n\,$.
Note, however, that the continuum part of the spectrum of ${\bf
H}_{\rm el}(R)$ can be implicitly included in this way as well, by
taking advantage of the box quantization procedure. For the sake
of clarity, let us also recall once again that the electron spin
variables have been suppressed in above formulas, although they
are implicitly accounted for.

The wavefunctions defined by Eq.~(\ref{SCHE_el}) depend only
parametrically on the nuclear coordinates
$(\vec{R}_c,R,\vartheta,\varphi)\,$. For each fixed nuclear
configuration $(\vec{R}_c,R,\vartheta,\varphi)\,$, the associated
collection of electronic states $\{ |\Phi_n(t)\ra \}$ forms a
complete orthonormal basis set covering an entire Hilbert space of
the electronic variables. The orthonormality and closure
properties are granted here for every time instant $t$, since the
time evolution according to the Schr\"{o}dinger equation,
Eq.~(\ref{SCHE_el}), is unitary, and since the field free
electronic eigenfunctions (\ref{evf_el_FF}) entering into the
initial condition (\ref{intcond}) constitute themselves a complete
orthonormal electronic basis set.

We note by passing that although the above initial condition
(\ref{intcond}) can be in principle used in the most general
context, it carries an especially physically illuminating
interpretation in the case when
\be
   \vec{A}(\vec{R}_c,t) \; = \; \vec{0} \mez , \mez \phi(\vec{R}_c,t) \; = \; 0
   \mez , \mez {\bf W}_{\rm el}(\vec{R}_c,R,\vartheta,\varphi,t)
   \; = \; {\bf 0} \mez \mez (t \leq t_0 ) \mez ;
\ee
valid for all the possible molecular positions $\vec{R}_c$
under study. If so, each stationary quantum state of the
considered molecule can be (of course for $t \leq t_0$ and in the
absence of avoided crossings) characterized within the framework
of the conventional Born-Oppenheimer approximation, as a product
of a specific electronic state $|\Phi_n^0\ra \exp[-(i/\hbar){\cal
E}_n^0(R)(t-t_0)]$ and an appropriate nuclear component. When the
field is switched on at $t > t_0\,$, it is natural to expect that
the relevant electronic state $n$ remains relatively well defined,
provided that its time evolution is understood in the sense of
equation (\ref{SCHE_el}). This idea stands behind our formulation
of the generalized time dependent Born-Oppenheimer separation
scheme which is elaborated in the coming subsection.

\subsection{ Generalized Born-Oppenheimer theory for time dependent Hamiltonians}

Following the spirit of the Born-Oppenheimer approach, we expand
the total molecular wavefunction as
\be \label{BO_expansion}
   \Psi(\vec{R}_c,R,\vartheta,\varphi,\vec{\tt r}^N\m,t) \; = \;
   \sum_n \, \chi_n(\vec{R}_c,R,\vartheta,\varphi,t) \;
   \Phi_n(\vec{\tt r}^N\m,t;\vec{R}_c,R,\vartheta,\varphi) \mez .
\ee
Substitution of an ansatz (\ref{BO_expansion}) into the time
dependent Schr\"{o}dinger equation (\ref{SCHE_2}) leads towards a
set of coupled differential equations for the as yet unknown
expansion coefficients $\chi_n\,$. That is,
\be \label{CC_SCHE}
   i\hbar \, \frac{\partial}{\partial t} \, \chi_n(\vec{R}_c,R,\vartheta,\varphi,t) \;
   = \; \sum_{n'} \, {\bf h}_{nn'}\n(t) \, \chi_{n'}\n(\vec{R}_c,R,\vartheta,\varphi,t)
   \mez ;
\ee
where the operator
\be
   {\bf h}_{nn'}\n(t) \; = \; {\bf h}_{nn'}^{(0)}\n(t) \; + \; {\bf h}_{nn'}^{(1)}\n(t)
   \; + \; {\bf h}_{nn'}^{(2)}\n(t)
\ee
is defined in terms of its components through the formulas
\be
   {\bf h}_{nn'}^{(0)}\n(t) \; = \; \delta_{nn'} \, \left[ \, - \, \frac{\hbar^2}{2 \,
   M} \, \Delta_{\vec{R}_c} \; - \; \frac{\hbar^2}{2 \, \mu_{{\cal AB}}} \, \frac{\partial^2}
   {\partial R^2} \; + \; \frac{\vec{\bf L}_{\vartheta\varphi}^2}{2 \, \mu_{{\cal AB}} \, R^2} \,
   \right] \mez ;
\ee
\vspace*{-0.50cm}
\begin{eqnarray} \label{h_nn'_1}
   {\bf h}_{nn'}^{(1)}\n(t) & = & - \, \frac{\hbar^2}{M} \; \la \Phi_n | \nabla_{\vec{R}_c}
   | \Phi_{n'} \ra_{\vec{\tt r}^N} \cdot \nabla_{\vec{R}_c} \; - \; \frac{\hbar^2}{\mu_{{\cal AB}}} \;
   \la \Phi_n | (\partial/\partial R) | \Phi_{n'} \ra_{\vec{\tt r}^N} \,
   \frac{\partial}{\partial R} \\
   & - & \frac{\hbar^2}{\mu_{{\cal AB}}R^2} \; \la \Phi_n | (\partial/\partial \vartheta)
   | \Phi_{n'} \ra_{\vec{\tt r}^N} \, \frac{\partial}{\partial \vartheta}
   \; - \; \frac{\hbar^2}{\mu_{{\cal AB}}R^2\sin^2\m\vartheta} \; \la \Phi_n |
   (\partial/\partial \varphi) | \Phi_{n'} \ra_{\vec{\tt r}^N} \,
   \frac{\partial}{\partial \varphi} \mez ; {}\nonumber
\end{eqnarray}
\vspace*{-0.50cm}
\begin{eqnarray} \label{h_nn'_2}
   {\bf h}_{nn'}^{(2)}\n(t) \; = \;
   \la \Phi_n | {\bf h}_{nn'}^{(2)}\n(t) | \Phi_{n'} \ra_{\vec{\tt r}^N}
   & = & - \, \frac{\hbar^2}{2M} \, \la \Phi_n | \Delta_{\vec{R}_c}
   | \Phi_{n'} \ra_{\vec{\tt r}^N} \; - \; \frac{\hbar^2}{2 \mu_{{\cal AB}}} \,
   \la \Phi_n | (\partial^2/\partial R^2) | \Phi_{n'} \ra_{\vec{\tt r}^N}
   {}\nonumber\\
   & - & \frac{\hbar^2}{2\mu_{{\cal AB}}R^2} \, \la \Phi_n | (\partial^2/\partial \vartheta^2)
   | \Phi_{n'} \ra_{\vec{\tt r}^N} \; - \; \frac{\hbar^2 \cos\vartheta}
   {2\mu_{{\cal AB}}R^2 \sin\vartheta} \; \la \Phi_n | (\partial/\partial \vartheta) | \Phi_{n'}
   \ra_{\vec{\tt r}^N} {}\nonumber\\
   & - & \frac{\hbar^2}{2\mu_{{\cal AB}}R^2\sin^2\m\vartheta} \, \la \Phi_n |
   (\partial^2/\partial \varphi^2) | \Phi_{n'} \ra_{\vec{\tt r}^N} \mez .
\end{eqnarray}
Here, for the sake of notational compactness, the
matrix elements over the electronic wavefunctions have been
expressed in a condensed fashion, such that e.g.
\be
   \la \Phi_n | (\partial/\partial R) | \Phi_{n'} \ra_{\vec{\tt r}^N}
   \; \equiv \; \int d^3{\tt r}^N \; \Phi_n^*(\vec{\tt r}^N\m,t;\vec{R}_c,R,\vartheta,\varphi) \,
   \frac{\partial}{\partial R} \, \Phi_{n'}\n(\vec{\tt r}^N\m,t;\vec{R}_c,R,\vartheta,\varphi)
   \mez ;
\ee
and similarly for the other quantities of this kind appearing
in Eqs.~(\ref{h_nn'_1})-(\ref{h_nn'_2}).

Relations (\ref{BO_expansion}) and (\ref{CC_SCHE}) are still
exact, as they constitute just a different equivalent
representation of the original problem (\ref{SCHE_2}). Now, let us
introduce the generalized adiabatic approximation, based upon
neglecting all the off-diagonal $(n \neq n')$ terms ${\bf
h}_{nn'}^{(1)}\n(t)$ and ${\bf h}_{nn'}^{(2)}\n(t)$ of the
Hamiltonian. If so, an index $n$ (used for labelling single
dynamical electronic basis vectors) becomes a good quantum number,
and equations (\ref{BO_expansion}) and (\ref{CC_SCHE}) are
simplified into \be \label{BOA}
   \Psi^{\rm BO}\n(\vec{R}_c,R,\vartheta,\varphi,\vec{\tt r}^N\m,t) \; = \;
   \chi_n^{\rm BO}\n(\vec{R}_c,R,\vartheta,\varphi,t) \;
   \Phi_n(\vec{\tt r}^N\m,t;\vec{R}_c,R,\vartheta,\varphi) \mez ;
\ee
and
\be \label{SCHE_BOA}
   i\hbar \, \frac{\partial}{\partial t} \, \chi_n^{\rm BO}\n(\vec{R}_c,R,\vartheta,\varphi,t) \;
   = \; {\bf h}_{nn}\n(t) \, \chi_{n}^{\rm BO}\n(\vec{R}_c,R,\vartheta,\varphi,t) \mez .
\ee
Before proceeding further, let us briefly touch a question
regarding the justification of the above approximative step. In
analogy to the conventional case of the time independent (field
free) Born-Oppenheimer approximation, it seems reasonable to
suggest the perturbation method as a well defined way how to
establish a criterion of validity for an adiabatic ansatz
(\ref{BOA}). If so, one might speculate that the adiabatic
approximation (\ref{BOA})-(\ref{SCHE_BOA}) breaks down in presence
of "near-degeneracies" whose enhancement or suppression can be
strongly influenced by an electromagnetic field. An explicit
elaboration of the just sketched ideas seems relatively
straightforward but lies beyond the scope of the present text.

\subsection{Time dependent electronic potential energy surfaces: The general
case}

Equation (\ref{SCHE_BOA}) is naturally interpreted as an effective
time dependent Schr\"{o}dinger equation which governs the nuclear
motions of the ${\cal AB}$ molecule in a given dynamical
electronic state $n$. It is convenient to carry out an additional
transformation which converts the associated effective Hamiltonian
${\bf h}_{nn}(t)$ into a more standard functional form
understandable as a sum of the kinetic and the potential energy
operators. To accomplish the mentioned task, one needs to
eliminate the Hamiltonian term ${\bf h}_{nn}^{(1)}(t)$ containing
the first order derivatives with respect to the nuclear
coordinates. An appropriate procedure for this is based upon
making a factorization
\be \label{factorization_BO}
   \chi_n^{\rm BO}\n(\vec{R}_c,R,\vartheta,\varphi,t) \; = \;
   e^{w_n(\vec{R}_c,R,\vartheta,\varphi,t)} \, \xi_n(\vec{R}_c,R,\vartheta,\varphi,t)
   \mez ;
\ee
where an exponential factor
$w_n(\vec{R}_c,R,\vartheta,\varphi,t)$ is chosen to satisfy a
system of the first order partial differential equations
\be
\label{pde_12}
   \nabla_{\vec{R}_c} \, w_n(\vec{R}_c,R,\vartheta,\varphi,t) \; =
   \; - \, \la \Phi_n | \nabla_{\vec{R}_c} | \Phi_n \ra_{\vec{\tt r}^N}
   \mez , \mez \partial \, w_n/\partial R \; = \;
   - \, \la \Phi_n | (\partial/\partial R) | \Phi_n \ra_{\vec{\tt r}^N}
\ee
\be \label{pde_34}
   \partial \, w_n/\partial \vartheta \; = \; - \,
   \la \Phi_n | (\partial/\partial \vartheta) | \Phi_n \ra_{\vec{\tt r}^N} \mez , \mez
   \partial \, w_n/\partial \varphi \; = \; - \,
   \la \Phi_n | (\partial/\partial \varphi) | \Phi_n \ra_{\vec{\tt r}^N} \mez ;
\ee
valid at every time instant $t\,$.

The question on general solvability of
Eqs.~(\ref{pde_12})-(\ref{pde_34}) is nontrivial and it is out of
the scope of this paper to discuss it. We shall assume for now
that the desired solution $w_n(\vec{R}_c,R,\vartheta,\varphi,t)$
exists and is unique up to a constant factor. This is of course
the case of a field free problem, where the time dependent
electronic wavefunctions depend only on $R$ and not on the other
nuclear coordinates $(\vec{R}_c,\vartheta,\varphi)\,$. We refer
also to other important special cases discussed in subsequent
subsections III.D and III.E, where an exact or at least a well
defined approximative solution can be shown to exist. In this
context one should note that, even if the set of
Eqs.~(\ref{pde_12})-(\ref{pde_34}) turns out to be generally not
solvable, an essential concept of the Born-Oppenheimer separation
represented by Eqs.~(\ref{BOA})-(\ref{SCHE_BOA}) remains
unaffected. The only controversy persists here on whether or not
one can formulate the general Born-Oppenheimer theory in terms of
the electronic potential energy surfaces. If not, one has to use
the above untransformed Born-Oppenheimer Hamiltonian ${\bf
h}_{nn}\n(t)$ which contains first order derivatives with respect
to the nuclear coordinates.

Having fixed the exponential factor
$w_n(\vec{R}_c,R,\vartheta,\varphi,t)$ according to
Eqs.~(\ref{pde_12})-(\ref{pde_34}), an equation of motion for the
wavefunction component $\xi_n(\vec{R}_c,R,\vartheta,\varphi,t)$ is
found to be
\be \label{SCHE_BO_V}
   i\hbar \, \frac{\partial}{\partial t} \, \xi_n(\vec{R}_c,R,\vartheta,\varphi,t) \;
   = \; {\bf h}_n^{\rm eff}\n(t) \; \xi_{n}(\vec{R}_c,R,\vartheta,\varphi,t) \mez ;
\ee where an effective translational-rotational-vibrational
Hamiltonian \be \label{h_n_eff}
   {\bf h}_n^{\rm eff}\n(t) \; = \;
   - \, \frac{\hbar^2}{2 M} \, \Delta_{\vec{R}_c} \; - \;
   \frac{\hbar^2}{2 \, \mu_{{\cal AB}}} \, \frac{\partial^2}{\partial R^2} \; + \;
   \frac{\vec{\bf L}_{\vartheta\varphi}^2}{2 \, \mu_{{\cal AB}} \, R^2} \; + \;
   V_n(\vec{R}_c,R,\vartheta,\varphi,t)
\ee
can be, indeed, understood as the kinetic energy operator of
the nuclei plus an electronic potential term. An explicit
expression for the mentioned potential is
\be \label{V_n_BO_tot}
   V_n(\vec{R}_c,R,\vartheta,\varphi,t) \; = \; - \, i\hbar \, \frac{\partial}{\partial t}
   \, w_n(\vec{R}_c,R,\vartheta,\varphi,t) \; + \; \tilde{V}_n(\vec{R}_c,R,\vartheta,\varphi,t)
   \mez ;
\ee
with the term
\begin{eqnarray} \label{V_n_BO_corr}
   \tilde{V}_n(\vec{R}_c,R,\vartheta,\varphi,t) & = & \Bigl[ \, {\bf h}_{nn}^{(0)}\n(t) \, + \,
   {\bf h}_{nn}^{(1)}\n(t) \, \Bigr] \, w_n(\vec{R}_c,R,\vartheta,\varphi,t) \; + \;
   {\bf h}_{nn}^{(2)}\n(t) {}\nonumber\\
   & - & \frac{\hbar^2}{2M} \, \left[ \nabla_{\vec{R}_c} w_n \, \right]^2
   \; - \; \frac{\hbar^2}{2\mu_{{\cal AB}}} \; \Bigl[ \, \partial w_n/\partial R \,
   \Bigr]^2 {}\nonumber\\
   & - & \frac{\hbar^2}{2\,\mu_{{\cal AB}}R^2} \; \Bigl[ \, \partial w_n/\partial \vartheta \, \Bigr]^2
   \; - \; \frac{\hbar^2}{2\,\mu_{{\cal AB}}R^2\sin^2\m\vartheta} \; \Bigl[ \, \partial w_n/\partial
   \varphi \, \Bigr]^2 \mez .
\end{eqnarray}
As a matter of fact, the dominant contribution to
$V_n(\vec{R}_c,R,\vartheta,\varphi,t)$ is given by the first part
of above equation (\ref{V_n_BO_tot}). Magnitude of the second term
(\ref{V_n_BO_corr}) can be generally considered to be small, due
to presence of inverse mass factors $M^{-1}$ and $\mu_{{\cal
AB}}^{-1}\,$. Similarly as in the usual textbook treatments of the
time independent Born-Oppenheimer separation \cite{BOA}, the
correction (\ref{V_n_BO_corr}) can be assigned to a difference
between the Born-Oppenheimer and the adiabatic approximations. We
shall neglect the term (\ref{V_n_BO_corr}) in our subsequent
considerations, taking just \be \label{V_n_BO}
   V_n(\vec{R}_c,R,\vartheta,\varphi,t) \; = \; - \, i\hbar \, \frac{\partial}{\partial t}
   \, w_n(\vec{R}_c,R,\vartheta,\varphi,t) \mez .
\ee

\subsection{ Electronic potential energy surfaces: The dc-field limit}

Let us investigate now a special case when the electromagnetic
field varies only very slowly in time. Such a situation is
referred as the dc (direct current) limit, and is encountered
whenever an external electrostatic field $\phi(\vec{r},t)$ is
turned on and off sufficiently slowly, while the central frequency
$\omega_L$ of the laser pulse $\vec{A}(\vec{r},t)$ acquires a
sufficiently small value (small with respect to the characteristic
timescale of the electronic motions - i.e.~practically even for
microwaves). Theoretical analysis of this problem is facilitated
by introducing a switching parameter $\eta(t)$ which modulates the
slow field variations according to a formal prescription
\be
\label{dc_fields}
   \vec{E}(\vec{r},t) \; = \; \vec{E}_{\rm dc}\n(\vec{r},\eta(t)) \mez , \mez
   d\eta(t)/dt \to 0 \mez .
\ee
For the sake of clarity, we quote in this context a simple
example of a low (microwave) frequency CW light
$\vec{E}(\vec{r},t) = \vec{E}_0 \cos(\omega_L
t-\vec{k}_L\cdot\vec{r})$ with $|\vec{k}_L| = \omega_L/c$ and
$\vec{E}_0 \cdot \vec{k}_L = 0\,$. Here the switching parameter
$\eta(t)$ can be defined just as $\omega_L t$ and is then
interpreted as an indicator of an instantaneous value of the
electric field strength.

Taking into account the property (\ref{dc_fields}) of the field,
it is clear that also the Hamiltonian interaction term
(\ref{W_el}) depends on time only through $\eta(t)\,$, and can be
thus denoted by an extra symbol
\be \label{W_el_dc}
   \hspace*{-0.50cm}
   {\bf W}_{\rm el}^{\rm ad}\n(\vec{R}_c,R,\vartheta,\varphi,\eta(t)) \; \equiv \;
   {\bf W}_{\rm el}(\vec{R}_c,R,\vartheta,\varphi,t) \; = \; - \, \Bigl[ \,
   {\cal M}(\vartheta,\varphi) \, \vec{D}_{{\cal AB}}^{BF}(R,\vartheta,\varphi,\vec{\tt r}^{\,N}) \, \Bigr]
   \cdot \vec{E}_{\rm dc}\n(\vec{R}_c,\eta(t)) \; \; .
\ee
If so, the dynamical electronic states (\ref{SCHE_el}) take an
explicit functional form predicted by the well known adiabatic
theorem \cite{Messiah}. It holds
\be \label{Phi_n_dc}
   \Phi_n(\vec{\tt r}^N\m,t;\vec{R}_c,R,\vartheta,\varphi) \; = \;
   \exp\m\left[ \, - \, \frac{i}{\hbar} \, \int_{t_0}^{t} \,
   {\cal E}_n^{\rm ad}\n(\vec{R}_c,R,\vartheta,\varphi,\eta(\tau)) \; d\tau \, \right] \,
   \Phi_n^{\rm ad}\n(\vec{\tt r}^N\m;\vec{R}_c,R,\vartheta,\varphi,\eta(t)) \mez ;
\ee
where the $\eta$-adiabatic electronic energies and
wavefunctions are defined by an eigenvalue problem
\be
\label{TI_SCHE_dc}
   \hspace*{-0.70cm}
   \Bigl[ \, {\bf H}_{\rm el}(R) \, + \, {\bf W}_{\rm el}^{\rm ad}\n(\vec{R}_c,R,\vartheta,\varphi,\eta) \, \Bigr]
   \, \Phi_n^{\rm ad}\n(\vec{\tt r}^N\m;\vec{R}_c,R,\vartheta,\varphi,\eta) \; = \;
   {\cal E}_n^{\rm ad}\n(\vec{R}_c,R,\vartheta,\varphi,\eta)
   \, \Phi_n^{\rm ad}\n(\vec{\tt r}^N\m;\vec{R}_c,R,\vartheta,\varphi,\eta) \; \; .
\ee
Since the electronic Hamiltonian of equation
(\ref{TI_SCHE_dc}) is real, the associated $\eta$-adiabatic
electronic eigenstates $\Phi_n^{\rm ad}\n(\vec{\tt
r}^N\m;\vec{R}_c,R,\vartheta,\varphi,\eta)$ can be also considered
as real quantities. Therefore, the matrix element
\be
   \la \Phi_n^{\rm ad} | (\partial/\partial R) | \Phi_n^{\rm ad} \ra_{\vec{\tt r}^N} \; = \; 0 \mez ;
\ee
and, consequently,
\be
   \la \Phi_n | (\partial/\partial R) | \Phi_n \ra_{\vec{\tt r}^N}
   \; = \; - \, \frac{i}{\hbar} \, \frac{\partial}{\partial R} \int_{t_0}^{t} \,
   {\cal E}_n^{\rm ad}\n(\vec{R}_c,R,\vartheta,\varphi,\eta(\tau)) \; d\tau
   \mez .
\ee
Analogical relations are valid also for the other matrix
elements appearing on right hand sides of
Eqs.~(\ref{pde_12})-(\ref{pde_34}). This shows that an appropriate
solution of the problem (\ref{pde_12})-(\ref{pde_34}) possesses
the form
\be \label{w_n_dc}
   w_n(\vec{R}_c,R,\vartheta,\varphi,t) \; = \; \frac{i}{\hbar} \, \int_{t_0}^{t} \,
   {\cal E}_n^{\rm ad}\n(\vec{R}_c,R,\vartheta,\varphi,\eta(\tau)) \; d\tau
   \mez ;
\ee
and the corresponding Born-Oppenheimer electronic potential
energy surface (\ref{V_n_BO}) reads as
\be \label{V_n_BO_dc}
   V_n(\vec{R}_c,R,\vartheta,\varphi,t) \; = \; {\cal E}_n^{\rm
   ad}\n(\vec{R}_c,R,\vartheta,\varphi,\eta(t)) \mez .
\ee
Equation (\ref{V_n_BO_dc}) displays of course an intuitively
expected result: {\it The calculated electronic potentials
coincide with those obtained within the conventional
Born-Oppenheimer approximation for the time independent static
field.}

\subsection{ Electronic potential energy surfaces: The ac-field
limit}

Here we investigate another special case when the electromagnetic
field oscillates rapidly and (quasi)periodically in time.
Mentioned situation is referred as the ac (alternating current)
limit, and is encountered whenever the studied molecule is exposed
to an adiabatically switched continuous wave UV-VIS-NIR laser, \be
\label{ac_field}
   \vec{E}(\vec{r},t) \; = \; \vec{E}_{\rm ac}^{}\n(\vec{r},\eta(t)) \;
   e^{+i \omega_L t}/2 \; + \; \vec{E}_{\rm ac}^{{}*}\n(\vec{r},\eta(t)) \;
   e^{-i \omega_L t}/2 \mez .
\ee
Here, symbol $\omega_L$ stands for the laser frequency, and
the field amplitude $\vec{E}_{\rm ac}^{}\n(\vec{r},\eta(t))$ is
allowed to depend very slowly (adiabatically) on a formally
introduced switching parameter $\eta(t)\,$. To avoid confusion, we
quote in this context a simple example of a CW-like Gaussian laser
pulse $\vec{E}_{\rm ac}^{}\n(\vec{r},\eta(t)) = \vec{E}_0 \;
e^{-\sigma t^2} \, e^{-i\vec{k}_L \cdot \vec{r}}$  with real
parameter $\sigma \to +0\,$, $|\vec{k}_L| = \omega_L/c$ and
$\vec{E}_0 \cdot \vec{k}_L = 0\,$. Here the switching parameter
$\eta(t)$ can be set to $e^{-\sigma t^2}$ and is then interpreted
as an indicator of an instantaneous value of an envelope of the
considered light pulse. For the sake of completeness, we note also
that the electrostatic potential $\phi(\vec{r},t)$ has been chosen
to be zero in the present example.


Taking into account the property (\ref{ac_field}) of the field, it
is clear that the corresponding Hamiltonian interaction term
(\ref{W_el}) depends on time only through $e^{\pm i \omega_L t}$
and $\eta(t)$, and can be thus denoted as
\begin{eqnarray}
\label{W_el_ac}
   {\bf W}_{\rm el}^{\rm F}\n(\vec{R}_c,R,\vartheta,\varphi,\eta(t),t) & \equiv &
   {\bf W}_{\rm el}(\vec{R}_c,R,\vartheta,\varphi,t) \; = {}\nonumber\\ & = & - \, \Bigl[ \,
   {\cal M}(\vartheta,\varphi) \, \vec{D}_{{\cal AB}}^{BF}(R,\vartheta,\varphi,\vec{\tt r}^{\,N}) \, \Bigr]
   \cdot \vec{E}_{\rm ac}^{}\n(\vec{r},\eta(t)) \; e^{+i \omega_L t}/2
   \; + \; {\rm c.c.}
\end{eqnarray}
If so, the dynamical electronic wavefunctions (\ref{SCHE_el}) take
an explicit functional form predicted by the adiabatic theorem for
the Floquet states \cite{Floquet}. It holds
\be \label{Phi_n_ac}
   \Phi_n(\vec{\tt r}^N\m,t;\vec{R}_c,R,\vartheta,\varphi) \; = \;
   \exp\m\left[ \, - \, \frac{i}{\hbar} \, \int_{t_0}^{t} \,
   {\cal E}_n^{\rm F}\n(\vec{R}_c,R,\vartheta,\varphi,\eta(\tau)) \; d\tau \, \right] \,
   \Phi_n^{\rm F}\n(\vec{\tt r}^N\m;\vec{R}_c,R,\vartheta,\varphi,\eta(t),t) \mez ;
\ee
where the $\eta$-adiabatic Floquet quasienergies and
eigenfunctions are defined by a generalized eigenvalue problem
$$
  \left[ \, {\bf H}_{\rm el}(R) \, + \, {\bf W}_{\rm el}^{\rm F}\n(\vec{R}_c,R,\vartheta,\varphi,\eta,t)
   \, - \, i\hbar \, (\partial/\partial t) \, \right]
   \, \Phi_n^{\rm F}\n(\vec{\tt r}^N\m;\vec{R}_c,R,\vartheta,\varphi,\eta,t) \; =
$$
\be \label{Floquet}
   = \; {\cal E}_n^{\rm F}\n(\vec{R}_c,R,\vartheta,\varphi,\eta)
   \; \Phi_n^{\rm F}\n(\vec{\tt r}^N\m;\vec{R}_c,R,\vartheta,\varphi,\eta,t) \mez .
\ee
In above equation (\ref{Floquet}), the time variable $t$ is
treated as an additional dynamical coordinate subjected to a
boundary condition
\be
   \Phi_n^{\rm F}\n(\vec{\tt r}^N\m;\vec{R}_c,R,\vartheta,\varphi,\eta,t) \; = \;
   \Phi_n^{\rm F}\n(\vec{\tt r}^N\m;\vec{R}_c,R,\vartheta,\varphi,\eta,t+T) \mez , \mez
   T \; = \; 2\pi/\omega_L \mez ;
\ee
in accordance with the spirit of the Floquet and $(t,t')$
theories \cite{Floquet_general}. The Floquet wavefunctions can be
expanded using the field free electronic basis set
(\ref{evf_el_FF}) into a sum
\be \label{Floquet_ef}
   \Phi_n^{\rm F}\n(\vec{\tt r}^N\m;\vec{R}_c,R,\vartheta,\varphi,\eta,t) \; = \;
   \sum_{n'} \sum_{m=-\infty}^{m=+\infty} \,
   C_{mn'}^n\n(\vec{R}_c,R,\vartheta,\varphi,\eta) \; \Phi_{n'}^0\n(\vec{\tt
   r}^N\m;R) \; e^{+i m \omega_L t} \mez ;
\ee
which is reduced just to a single term $\Phi_n^0\n(\vec{\tt
r}^N\m;R)$ as soon as the field amplitude is turned off. Written
mathematically,
\be
   C_{mn'}^n\n(\vec{R}_c,R,\vartheta,\varphi,\eta_0) \; = \;
   \delta_{nn'} \, \delta_{m0} \mez {\rm for} \mez
   \vec{E}_{\rm ac}^{}\n(\vec{r},\eta_0) \; = \; \vec{0} \mez .
\ee
Similarly, also
\be
   {\cal E}_n^{\rm F}\n(\vec{R}_c,R,\vartheta,\varphi,\eta_0) \; =
   \; {\cal E}_n^0 \mez .
\ee
Strictly speaking, the above outlined formulation of the
Floquet theory is physically adequate only in the weak field
regime where the associated Floquet states resemble the properties
of the bound states. For strong fields, where the field induced
ionization phenomenon becomes important, the problem must be
addressed within the framework of the non-Hermitian quantum
mechanics, with different types of complex scaling transformations
being employed to yield {\it complex} quasienergies
\cite{NM-report}. In such a case an imaginary part of the
quasienergy corresponds to an inverse lifetime of the associated
metastable electronic Floquet state. Further details regarding the
Floquet theory can be found in Refs.~\cite{Floquet_general}.

Using an adiabatic ansatz (\ref{Phi_n_ac}) we find that the matrix
element
\be
   \la \Phi_n | (\partial/\partial R) | \Phi_n \ra_{\vec{\tt r}^N}
   \; = \; - \, \frac{i}{\hbar} \, \frac{\partial}{\partial R} \int_{t_0}^{t} \,
   {\cal E}_n^{\rm F}\n(\vec{R}_c,R,\vartheta,\varphi,\eta(\tau)) \; d\tau
   \; + \; \la \Phi_n^{\rm F} | (\partial/\partial R) | \Phi_n^{\rm F} \ra_{\vec{\tt r}^N}
   \mez .
\ee
Analogical relations are valid also for the other matrix
elements appearing on right hand sides of equations
(\ref{pde_12})-(\ref{pde_34}). Since the system of differential
equations (\ref{pde_12})-(\ref{pde_34}) is linear and homogeneous,
an appropriate solution should possess the form
\be \label{w_n_ac}
   w_n(\vec{R}_c,R,\vartheta,\varphi,t) \; = \; \frac{i}{\hbar} \, \int_{t_0}^{t} \,
   {\cal E}_n^{\rm F}\n(\vec{R}_c,R,\vartheta,\varphi,\eta(\tau)) \; d\tau
   \; + \; \tilde{w}_n(\vec{R}_c,R,\vartheta,\varphi,t) \mez ;
\ee
where the quantity
$\tilde{w}_n(\vec{R}_c,R,\vartheta,\varphi,t)$ is assumed to
satisfy a system of the first order partial differential equations
\be \label{pde_12_F}
   \nabla_{\vec{R}_c} \, \tilde{w}_n(\vec{R}_c,R,\vartheta,\varphi,t) \; =
   \; - \, \la \Phi_n^{\rm F} | \nabla_{\vec{R}_c} | \Phi_n^{\rm F} \ra_{\vec{\tt r}^N}
   \mez , \mez \partial \, \tilde{w}_n/\partial R \; = \;
   - \, \la \Phi_n^{\rm F} | (\partial/\partial R) | \Phi_n^{\rm F} \ra_{\vec{\tt r}^N}
\ee
\be \label{pde_34_F}
   \partial \, \tilde{w}_n/\partial \vartheta \; = \; - \,
   \la \Phi_n^{\rm F} | (\partial/\partial \vartheta) | \Phi_n^{\rm F} \ra_{\vec{\tt r}^N} \mez , \mez
   \partial \, \tilde{w}_n/\partial \varphi \; = \; - \,
   \la \Phi_n^{\rm F} | (\partial/\partial \varphi) | \Phi_n^{\rm F} \ra_{\vec{\tt r}^N} \mez .
\ee
The question on solvability of
Eqs.~(\ref{pde_12_F})-(\ref{pde_34_F}) does not seem to be less
difficult than in the case of Eqs.~(\ref{pde_12})-(\ref{pde_34}).
For this reason, we prefer to carry out the phase transformation
(\ref{factorization_BO}) using only the factor
\be
\label{w_n_ac_approx}
   w_n(\vec{R}_c,R,\vartheta,\varphi,t) \; = \; \frac{i}{\hbar} \, \int_{t_0}^{t} \,
   {\cal E}_n^{\rm F}\n(\vec{R}_c,R,\vartheta,\varphi,\eta(\tau)) \; d\tau
   \mez .
\ee
Since the function (\ref{w_n_ac_approx}) does not represent an
exact solution of the problem (\ref{pde_12})-(\ref{pde_34}),
the first derivatives of the hamiltonian term ${\bf
h}_{nn}^{(1)}\n(t)$ (see Eq.~(\ref{h_nn'_1}) for $n=n'$) are not
completely eliminated. Instead, ${\bf h}_{nn}^{(1)}\n(t)$ is
transformed into
\begin{eqnarray} \label{h_nn_1_F}
   {\bf h}_{nn}^{\rm (F)}\n(t) & = & - \, \frac{\hbar^2}{M} \; \la \Phi_n^{\rm F} | \nabla_{\vec{R}_c}
   | \Phi_{n}^{\rm F} \ra_{\vec{\tt r}^N} \cdot \nabla_{\vec{R}_c} \; - \; \frac{\hbar^2}{\mu_{{\cal AB}}} \;
   \la \Phi_n^{\rm F} | (\partial/\partial R) | \Phi_{n}^{\rm F} \ra_{\vec{\tt r}^N} \,
   \frac{\partial}{\partial R} \\
   & - & \frac{\hbar^2}{\mu_{{\cal AB}}R^2} \; \la \Phi_n^{\rm F} | (\partial/\partial \vartheta)
   | \Phi_{n}^{\rm F} \ra_{\vec{\tt r}^N} \, \frac{\partial}{\partial \vartheta}
   \; - \; \frac{\hbar^2}{\mu_{{\cal AB}}R^2\sin^2\m\vartheta} \; \la \Phi_n^{\rm F} |
   (\partial/\partial \varphi) | \Phi_{n}^{\rm F} \ra_{\vec{\tt r}^N} \,
   \frac{\partial}{\partial \varphi} \mez . {}\nonumber
\end{eqnarray}
Due to presence of the above first derivative term in the effective nuclear hamiltonian,
the obtained electronic potential energy surface
\be \label{V_n_BO_ac}
   V_n(\vec{R}_c,R,\vartheta,\varphi,t) \; = \; {\cal E}_n^{\rm
   F}\n(\vec{R}_c,R,\vartheta,\varphi,\eta(t))
\ee
accounts only partially for the underlying Born-Oppenheimer
nuclear dynamics. Nevertheless, as being explained in the next
paragraph, the transformation (\ref{factorization_BO}) with the
phase factor (\ref{w_n_ac_approx}) still proves to be an important
step which offers a lot of physical insight.

The nature of the Floquet wavefunctions (\ref{Floquet_ef}) reveals
that all the diagonal matrix elements contained in the formula
(\ref{h_nn_1_F}) depend on time solely through the oscillating
factors $e^{\pm i m \omega_L t}$ ($m$ nonzero integer). Validity
of this statement can be most directly demonstrated through an
evaluation of the time averages
\be
   \int_{t}^{t+T} \la \Phi_n^{\rm F}(\tau) | (\partial/\partial \Omega)
   | \Phi_{n}^{\rm F}(\tau) \ra_{\vec{\tt r}^N} \; d\tau \; = \; 0 \mez ,
   \mez \Omega \; = \; (X,Y,Z,R,\vartheta,\varphi) \mez ;
\ee
by taking advantage of the Feynman-Hellman theory adapted for
the Floquet states \cite{Floquet}. Due to presence of the just
discussed oscillatory factors, the quantity (\ref{V_n_BO_ac}) can
be regarded as a physically well justified Born-Oppenheimer
potential term in the ac-field limit. This holds true as long as
the duration $T=2\pi/\omega_L$ of one optical cycle remains much smaller than the
characteristic timescales of the nuclear motions, so that the
oscillating first derivative contribution (\ref{h_nn_1_F}) is
irrelevant. On the other hand, the correction (\ref{h_nn_1_F})
becomes increasingly important as $\omega_L$ decreases and
approaches the dc-field limit. One can see it immediately also
from the fact that the above ac-field potential (\ref{V_n_BO_ac})
depends on time only through an {\it envelope} of the light pulse,
while the correct dc-field formula (\ref{V_n_BO_dc}) is defined in
terms of an {\it instantaneous} field strength and includes thus
the field oscillations $e^{\pm i \omega_L t}$. The coming Section
IV provides a more explicit comparison of the Born-Oppenheimer
potentials in the dc-field and ac-field limits.


\section{ Diatomic molecule in an electromagnetic field:\\
Electronic potential energy surfaces in the weak field regime}

\subsection{ Application of the perturbation theory in the dc-field limit }

Provided that the used field strength is sufficiently weak, the
corresponding Born-Oppenheimer electronic eigenenergies and
wavefunctions defined by equation (\ref{TI_SCHE_dc}) are only
slightly different from their field free counterparts. If so, the
framework of perturbation theory offers a straightforward method
for resolving the mentioned Born-Oppenheimer electronic problem,
and leads towards explicit analytic formulas for the field induced
corrections of the associated potential energy surfaces, ${\cal
E}_n^{\rm ad}\n(\vec{R}_c,R,\vartheta,\varphi,\eta) - {\cal
E}_n^0\n(R)\,$. Clearly, the field strength is considered here to
be the perturbation expansion parameter, while the solutions of
the field free eigenproblem (\ref{evf_el_FF}) are taken as an
unperturbed reference.

Before presenting the details of the perturbation approach, we
would like to emphasize that this is not the only practical method
for evaluation of the desired dc-field Born-Oppenheimer electronic
potential energy surfaces. An alternative possibility is to use
direct ({\it ab initio}) numerical solution of equation
(\ref{TI_SCHE_dc}), which is of course much more demanding from
computational point of view, but remains appropriate even in the
case of strong dc-fields.

Perturbation expansion for the energy eigenvalue of equation
(\ref{TI_SCHE_dc}) possesses the form \be
   {\cal E}_n^{\rm ad}\n(\vec{R}_c,R,\vartheta,\varphi,\eta) \; =
   \; {\cal E}_n^0(R) \; + \; {\cal E}_n^{\rm ad,1}\n(\vec{R}_c,R,\vartheta,\varphi,\eta)
   \; + \; {\cal E}_n^{\rm ad,2}\n(\vec{R}_c,R,\vartheta,\varphi,\eta)
   \; + \; \cdots \mez ;
\ee
where the first order correction
\be
   {\cal E}_n^{\rm ad,1}\n(\vec{R}_c,R,\vartheta,\varphi,\eta) \;
   = \; \la \Phi_n^0 | {\bf W}_{\rm el}^{\rm ad}\n(\vec{R}_c,R,\vartheta,\varphi,\eta)
   | \Phi_n^0 \ra_{\vec{\tt r}^N} \mez ;
\ee
and the second order term
\be
   {\cal E}_n^{\rm ad,2}\n(\vec{R}_c,R,\vartheta,\varphi,\eta) \;
   = \; \sum_{n' \neq n} \, \frac{ \Bigl| \, \la \Phi_n^0 |
   {\bf W}_{\rm el}^{\rm ad}\n(\vec{R}_c,R,\vartheta,\varphi,\eta)
   | \Phi_{n'}^0\m \ra_{\vec{\tt r}^N} \, \Bigr|^2 }
   {{\cal E}_n^0(R)-{\cal E}_{n'}^0\n(R)} \mez .
\ee Substitution of an expression (\ref{W_el_dc}) yields
consequently more explicit results \be \label{E_dc_PT1}
   {\cal E}_n^{\rm ad,1}\n(\vec{R}_c,R,\vartheta,\varphi,\eta) \;
   = \; - \, \left[ \, {\cal M}(\vartheta,\varphi) \, \la \Phi_n^0 | \vec{D}_{{\cal AB}}^{BF}(R,\vartheta,\varphi,\vec{\tt r}^{\,N})
   | \Phi_n^0 \ra_{\vec{\tt r}^N} \, \right] \cdot \vec{E}_{\rm dc}(\vec{R}_c,\eta) \mez ;
\ee and \be \label{E_dc_PT2}
   {\cal E}_n^{\rm ad,2}\n(\vec{R}_c,R,\vartheta,\varphi,\eta) \; = \; \sum_{n' \neq n} \,
   \frac{ \biggl| \, \left[ \, {\cal M}(\vartheta,\varphi) \, \la \Phi_n^0 | e \sum_{j=1}^{N}
   \vec{\tt r}_j | \Phi_{n'}^0 \m \ra_{\vec{\tt r}^N} \, \right] \cdot
   \vec{E}_{\rm dc}(\vec{R}_c,\eta) \, \biggr|^2 }{ {\cal E}_n^0(R)-{\cal E}_{n'}^0\n(R) } \mez .
\ee
Relation (\ref{E_dc_PT1}) determines of course an energy of an
interaction between the molecular permanent dipole moment
(associated with given electronic state $n$) and the total electric
field (arising both due to the scalar and the vector potentials). We
refer in this context to our discussion of the dipole term
$\vec{D}_{\cal AB}$ carried out at the end of subsection II.C. In
passing we note that the just mentioned kind of dipole interaction
constitutes a conceptual basis of the traditional rovibrational
spectroscopy (we recall that the dc-field limit is appropriate in
the microwave and the far infrared spectral domains).

Physical contents of equation (\ref{E_dc_PT2}) becomes apparent
after rewriting it into an equivalent fashion
\be
\label{E_dc_PT2_pol}
   {\cal E}_n^{\rm ad,2}\n(\vec{R}_c,R,\vartheta,\varphi,\eta) \; = \; - \, \frac{1}{2} \;
   \left[ \, {\cal M}(\vartheta,\varphi) \, \vec{E}_{\rm dc}^T(\vec{R}_c,\eta) \, \right]
   \cdot \dm{\alpha}_n\m\n(R) \cdot \left[ \, {\cal M}^T\n(\vartheta,\varphi) \, \vec{E}_{\rm
   dc}(\vec{R}_c,\eta) \, \right] \mez ;
\ee
where the symbol
\be \label{pol_dc_gen}
   \dm{\alpha}_n\m\n(R) \; = \; 2 \; \sum_{n' \neq n} \, \frac{ \la \Phi_n^0 | e \sum_{j=1}^{N}
   \vec{\tt r}_j | \Phi_{n'}^0 \m \ra_{\vec{\tt r}^N} \m \la \Phi_{n'}^0 | e \sum_{j=1}^{N}
   \vec{\tt r}_j | \Phi_n^0 \ra}{ {\cal E}_{n'}^0\n(R)-{\cal E}_n^0(R) }
\ee
is recognized as a tensor of static polarizability associated
with the $n$-th electronic state of the ${\cal AB}$ molecule. Since
the quantity $\dm{\alpha}_n\m\n(R)$ is expressed with respect to the
body fixed frame defined within subsection II.E, its off-diagonal
matrix elements necessarily vanish provided that the relevant
electronic state $|\Phi_n^0\ra$ is of $\Sigma$ type. (This is the
case of the ground electronic states of the vast majority of
diatomic molecules.) The situation is a bit more complicated in the
case when the electronic states of types $\Pi$, $\Delta$, etc.~enter
into the play. We shall omit all the details for the sake of
simplicity, and consider from now on just the $\Sigma$ type
electronic ground state $|\Phi_g^0\ra\,$. Symmetry considerations
dictate that
\be \label{alpha_dc_diag}
   \alpha_g^{xx}\n(R) \; = \; \alpha_g^{yy}\n(R) \; \equiv \; \alpha_g^{\perp}\n(R)
   \; \; , \; \; \alpha_g^{zz}\n(R) \; \equiv \; \alpha_g^{\parallel}\n(R) \; \; , \; \;
   \alpha_g^{xy}\n(R) \; = \; \alpha_g^{xz}\n(R) \; = \; \alpha_g^{yz}\n(R)
   \; = \; 0 \mez .
\ee Moreover, it is clear that the $x$ and $y$ components of the
dipole moment matrix element $\la \Phi_g^0 | \vec{D}_{{\cal
AB}}^{BF}(R,\vartheta,\varphi,\vec{\tt r}^{\,N}) | \Phi_g^0
\ra_{\vec{\tt r}^N}$ appearing in the first order correction
(\ref{E_dc_PT1}) are zero. In order to fully exploit these useful
properties, we assume without any loss of generality that the only
nonzero component of the dc electric field is pointing along the
space fixed $z$-axis, i.e. \be \label{E_dc_zsf}
   \vec{E}_{\rm dc}\n(\vec{R}_c,\eta) \; = \; \Bigl[ \, 0 , 0 ,
   E_{\rm dc}\n(\vec{R}_c,\eta) \, \Bigr] \mez .
\ee
Using equations (\ref{M_matrix}), (\ref{alpha_dc_diag}) and
(\ref{E_dc_zsf}), we simplify the formulas (\ref{E_dc_PT1}) and
(\ref{E_dc_PT2_pol}) into
\be \label{E_dc_PT1_simple}
   {\cal E}_g^{\rm ad,1}\n(\vec{R}_c,R,\vartheta,\varphi,\eta) \; = \; - \;
   E_{\rm dc}\n(\vec{R}_c,\eta) \, \cos\vartheta \,
   \la \Phi_g^0 | D_{{\cal AB}}^{BF,z}(R,\vartheta,\varphi,\vec{\tt r}^{\,N}) | \Phi_g^0 \ra_{\vec{\tt r}^N} \mez ;
\ee and \be \label{E_dc_PT2_pol_simple}
   {\cal E}_g^{\rm ad,2}\n(\vec{R}_c,R,\vartheta,\varphi,\eta) \; = \; - \, \frac{1}{2} \;
   E_{\rm dc}^2\n(\vec{R}_c,\eta) \, \left[ \, \alpha_g^\parallel\n(R) \, \cos^2\vartheta \; + \;
   \alpha_g^\perp\n(R) \, \sin^2\vartheta\, \right] \mez .
\ee
Having derived these two important expressions, we conclude
our discussion by writing down the final result for the associated
translational/rotational/vibrational Hamiltonian (\ref{h_n_eff})
assigned to the ground electronic state diatomic molecule ${\cal
AB}$ interacting with a weak external dc electromagnetic field. It
holds
\begin{eqnarray} \label{h_g_dc_final}
   {\bf h}_g^{\rm eff}\n(t) & = & - \, \frac{\hbar^2}{2 M} \, \Delta_{\vec{R}_c} \; - \;
   \frac{\hbar^2}{2 \, \mu_{{\cal AB}}} \, \frac{\partial^2}{\partial R^2} \; + \;
   \frac{\vec{\bf L}_{\vartheta\varphi}^2}{2 \, \mu_{{\cal AB}} \, R^2} \; - \;
   E_{\rm dc}\n(\vec{R}_c,\eta(t)) \, \cos\vartheta \,
   \la \Phi_g^0 | D_{{\cal AB}}^{BF,z}(R,\vartheta,\varphi,\vec{\tt r}^{\,N}) | \Phi_g^0 \ra_{\vec{\tt r}^N}
   {}\nonumber \\
   & - & \frac{1}{2} \;
   E_{\rm dc}^2\n(\vec{R}_c,\eta(t)) \, \left[ \, \alpha_g^\parallel\n(R) \, \cos^2\vartheta
   \; + \; \alpha_g^\perp\n(R) \, \sin^2\vartheta\, \right] \mez .
\end{eqnarray}
Note that the dc-field enters here through its {\it instantaneous}
time dependent strength $E_{\rm dc}(\vec{R}_c,\eta(t))\,$. As being
shown in the next subsection, this finding is in a sharp contrast to
the situation when an ac-field is present.

\subsection{Application of the perturbation theory in the ac-field limit}

The framework of the perturbation theory can be exploited as well in
the case of weak ac-fields. Perturbative treatment of the
Born-Oppenheimer-Floquet electronic problem (\ref{Floquet}) leads
towards explicit analytic formulas for the field induced corrections
of the associated potential energy surfaces, ${\cal E}_n^{\rm
F}\n(\vec{R}_c,R,\vartheta,\varphi,\eta) - {\cal E}_n^0\n(R)\,$.
Similarly as in the above dc-field case, the field strength is
considered to be the perturbation expansion parameter, while the
solutions of the field free eigenproblem (\ref{evf_el_FF}) are taken
as an unperturbed reference. In passing we note that, since the time
variable $t$ is within the Floquet formalism treated as an
additional dynamical coordinate, we can still rely here on a
(properly modified) time independent perturbation theory. Instead of
supplying a more detailed description of such a Floquet type time
independent perturbation approach, we refer to \cite{Floquet} and
\cite{Floquet_PT} for well elaborated examples.

Similarly as in above subsection IV.A, we would like to emphasize
that the perturbation approach is not the only practical method for
evaluation of the desired ac-field Born-Oppenheimer electronic
potential energy surfaces. An alternative possibility is to use
direct ({\it ab initio}) numerical solution of the Floquet
eigenproblem (\ref{Floquet}). Such a task is of course much more
demanding from the computational point of view, but remains
appropriate even in the case of strong ac-fields.

Perturbation expansion for the quasienergy eigenvalue of equation
(\ref{Floquet}) possesses the form \be
   {\cal E}_n^{\rm F}\n(\vec{R}_c,R,\vartheta,\varphi,\eta) \; =
   \; {\cal E}_n^0(R) \; + \; {\cal E}_n^{\rm F,1}\n(\vec{R}_c,R,\vartheta,\varphi,\eta)
   \; + \; {\cal E}_n^{\rm F,2}\n(\vec{R}_c,R,\vartheta,\varphi,\eta)
   \; + \; \cdots \mez ;
\ee where the first order correction ${\cal E}_n^{\rm
F,1}\n(\vec{R}_c,R,\vartheta,\varphi,\eta)$ vanishes due to time
averaging over the ac field oscillations, while the second order
term
\begin{eqnarray} \label{E_ac_PT2}
   {\cal E}_n^{\rm F,2}\n(\vec{R}_c,R,\vartheta,\varphi,\eta) & = & \sum_{n' \neq n} \,
   \left| \, \Bigl[ \, {\cal M}(\vartheta,\varphi) \, \la \Phi_n^0 | e \sum_{j=1}^{N}
   \vec{\tt r}_j | \Phi_{n'}^0 \m \ra_{\vec{\tt r}^N} \, \Bigr] \cdot
   \vec{E}_{\rm ac}^\perp(\vec{R}_c,\eta)/2 \; \right|^2 {}\nonumber\\
   & \times & \left\{ \, \frac{1}{ {\cal E}_n^0(R)-{\cal E}_{n'}^0\n(R)+\hbar\omega_L } \, + \,
   \frac{1}{ {\cal E}_n^0(R)-{\cal E}_{n'}^0\n(R)-\hbar\omega_L } \, \right\} \mez .
\end{eqnarray}
Physical contents of equation (\ref{E_ac_PT2}) becomes apparent
after rewriting it into an equivalent fashion \be
\label{E_ac_PT2_pol}
   {\cal E}_n^{\rm F,2}\n(\vec{R}_c,R,\vartheta,\varphi,\eta) \; = \; - \, \left[ \,
   {\cal M}(\vartheta,\varphi) \, \vec{E}_{\rm ac}^{{} T}\n(\vec{R}_c,\eta)/2 \, \right]
   \cdot \dm{\alpha}_n\n(R,\omega_L) \cdot \left[ \, {\cal M}^T\n(\vartheta,\varphi) \,
   \vec{E}_{\rm ac}^{{}}\n(\vec{R}_c,\eta)/2 \, \right] \mez ;
\ee
where the quantity
\be \label{pol_ac_gen}
   \dm{\alpha}_n\n(R,\omega_L) \; = \; \frac{2}{\hbar} \; \sum_{n' \neq n} \,
   \frac{ \la \Phi_n^0 | e \sum_{j=1}^{N} \vec{\tt r}_j | \Phi_{n'}^0 \m \ra_{\vec{\tt r}^N} \m
   \la \Phi_{n'}^0 | e \sum_{j=1}^{N} \vec{\tt r}_j | \Phi_n^0 \ra \; \omega_{nn'}\n(R) }
   { \omega_L^2 \, - \, \omega_{nn'}^2\n(R) }
\ee
is recognized as a tensor of dynamical (frequency dependent)
polarizability associated with the $n$-th electronic state of the
${\cal AB}$ molecule, and an auxiliary symbol
\be
   \omega_{nn'}\n(R) \; = \; \Bigl[ \, {\cal E}_n^0(R)-{\cal E}_{n'}^0\n(R) \, \Bigr]/\hbar
   \mez .
\ee
Since the tensor $\dm{\alpha}_n\m\n(R,\omega_L)$ is expressed
with respect to the body fixed frame, its off-diagonal matrix
elements necessarily vanish provided that we restrict ourselves on
the case of the ground electronic state $|\Phi_n^0\ra \equiv
|\Phi_g^0\ra$ which is assumed to be of $\Sigma$ type. Symmetry
considerations dictate then that
\be \label{alpha_ac_diag_1}
   \alpha_g^{xx}\n(R,\omega_L) \; = \; \alpha_g^{yy}\n(R,\omega_L) \; \equiv \;
   \alpha_g^{\perp}\n(R,\omega_L) \mez , \mez \alpha_g^{zz}\n(R,\omega_L) \; \equiv \;
   \alpha_g^{\parallel}\n(R,\omega_L) \mez ;
\ee
and
\be \label{alpha_ac_diag_2}
   \alpha_g^{xy}\n(R,\omega_L) \; = \; \alpha_g^{xz}\n(R,\omega_L) \; = \;
   \alpha_g^{yz}\n(R,\omega_L) \; = \; 0 \mez .
\ee
In order to fully exploit these useful properties, we assume
without any loss of generality that the only nonzero component of
the considered ac electric field is pointing along the space fixed
$z$-axis, i.e.,
\be \label{E_ac_zsf}
   \vec{E}_{\rm ac}^{}\n(\vec{R}_c,\eta) \; = \; \Bigl[ \, 0 , 0 ,
   E_{\rm ac}\n(\vec{R}_c,\eta) \, \Bigr] \mez .
\ee
Using Eqs.~(\ref{M_matrix}), (\ref{alpha_ac_diag_1}) and
(\ref{alpha_ac_diag_2}), we simplify the formula
(\ref{E_ac_PT2_pol}) into
\be \label{E_ac_PT2_pol_simple}
   {\cal E}_g^{\rm F,2}\n(\vec{R}_c,R,\vartheta,\varphi,\eta) \; = \; - \, \frac{1}{4} \;
   E_{\rm ac}^2\n(\vec{R}_c,\eta) \, \left[ \, \alpha_g^\parallel\n(R,\omega_L) \,
   \cos^2\vartheta \; + \; \alpha_g^\perp\n(R,\omega_L) \, \sin^2\vartheta\, \right] \mez .
\ee
Note that we encounter here a multiplicative factor $(1/4)$
contrary to the dc-field expression (\ref{E_dc_PT2_pol_simple})
where a factor $(1/2)$ appears instead. Let us write down now the
final result for the associated
translational/rotational/vibrational Hamiltonian (\ref{h_n_eff})
assigned to the ground electronic state diatomic molecule ${\cal
AB}$ interacting with a weak external ac electromagnetic field. It
holds
\begin{eqnarray} \label{h_g_ac_final}
   {\bf h}_g^{\rm eff}\n(t) & = & - \, \frac{\hbar^2}{2 M} \, \Delta_{\vec{R}_c} \; - \;
   \frac{\hbar^2}{2 \, \mu_{{\cal AB}}} \, \frac{\partial^2}{\partial R^2} \; + \;
   \frac{\vec{\bf L}_{\vartheta\varphi}^2}{2 \, \mu_{{\cal AB}} \, R^2} \\
   & - & \frac{1}{4} \; E_{\rm ac}^2\n(\vec{R}_c,\eta(t)) \, \left[ \,
   \alpha_g^\parallel\n(R,\omega_L) \, \cos^2\vartheta \; + \; \alpha_g^\perp\n(R,\omega_L) \,
   \sin^2\vartheta\, \right] \mez . {}\nonumber
\end{eqnarray}
Note that the ac-field enters here solely through its {\it
amplitude} $E_{\rm ac}\n(\vec{R}_c,\eta(t))\,$, while the rapid
field oscillations $e^{\pm i \omega_L t}$ do not appear. This
behavior is in a sharp contrast to the case of dc-field analyzed
within the previous subsection, see equation (\ref{h_g_dc_final}).

Our discussion concludes by pointing out one additional remark. The
perturbational analysis performed within this subsection has been
based upon the length gauge Hamiltonian defined by formulas
(\ref{H_full_simple})-(\ref{W_el}). Another option might be to
develop a perturbational expansion employing the momentum gauge
Hamiltonian (\ref{H_sf_cr_final}), of course after having converted
it properly into the body fixed frame following similar approach as
in above subsections II.D-II.E. Interestingly, the results obtained
within the length gauge and the momentum gauge perturbation theory
possess somewhat different functional forms. Our calculations reveal
that the momentum gauge expression for ${\cal E}_n^{\rm
F,2}\n(\vec{R}_c,R,\vartheta,\varphi,\eta)$ is still given by
relation (\ref{E_ac_PT2_pol}), with the dynamical polarizability
tensor being however redefined as
\be \label{pol_ac_gen_mg}
   \dm{\alpha}_n\n(R,\omega_L) \; = \; \frac{2}{\hbar} \; \sum_{n' \neq n} \,
   \frac{ \la \Phi_n^0 | e \sum_{j=1}^{N} \vec{\tt r}_j | \Phi_{n'}^0 \m \ra_{\vec{\tt r}^N} \m
   \la \Phi_{n'}^0 | e \sum_{j=1}^{N} \vec{\tt r}_j | \Phi_n^0 \ra \; \omega_{nn'}\n(R) }
   { \omega_L^2 \, - \, \omega_{nn'}^2\n(R) } \, \frac{\omega_{nn'}^2\n(R)}{\omega_L^2}
   \mez .
\ee
The only difference between formulas (\ref{pol_ac_gen}) and
(\ref{pol_ac_gen_mg}) consists in the presence of an extra factor
$(\omega_{nn'}\n(R)/\omega_L)^2\,$. Due to this extra factor, one
would be tempted to argue that equations (\ref{pol_ac_gen}) and
(\ref{pol_ac_gen_mg}) provide different results for the
polarizability. Even a direct numerical calculation using a finite
basis set of the field free molecular electronic states $\{ |
\Phi_{n'}^0 \ra \}$ shows that the predictions of the formulas
(\ref{pol_ac_gen}) and (\ref{pol_ac_gen_mg}) are different.
Especially pronounced discrepancies are found in the case when
$\omega_L$ is far off resonant from any electronic transition $|
\Phi_n^0 \ra \to | \Phi_{n'}^0\n\ra\,$. Mentioned observations might
seem to indicate inconsistencies, since the passage from the
momentum gauge to the length gauge is facilitated through an unitary
transformation, and the quantum mechanical perturbation theory is
known to be gauge invariant. A detailed theoretical analysis of the
above sketched problem is postponed into another paper \cite{gipt}.
Here we just state without proof that both expressions
(\ref{pol_ac_gen}) and (\ref{pol_ac_gen_mg}) turn out to be
identical, {\it provided that one properly accounts for a complete
set of all the electronic states, including also the highly excited
electronic continuum}. Use of a truncated basis set in equation
(\ref{pol_ac_gen_mg}) is not always justified and can provide an
absolutely misleading outcome. This example should serve as a
warning against an uncautious use of an incomplete electronic basis
set (e.g.~within the widely considered two level approximation) when
studying the field induced atomic/molecular properties in the
momentum gauge.


\section{Concluding remarks}

In the present paper, we have studied general problem of the
translational/rotational/vibra- tional/electronic dynamics of a
diatomic molecule exposed to an interaction with an arbitrary
external electromagnetic field. We have derived an appropriate body
fixed frame Hamiltonian (\ref{H_full_simple}), and introduced the
concept of the time dependent Born-Oppenheimer approximation. An
interesting open question on general existence of the time dependent
Born-Oppenheimer electronic potential energy surfaces has been
raised. Finally, we have derived an effective
translational/rotational/vibrational Hamiltonian
(\ref{h_g_dc_final}) resp.~(\ref{h_g_ac_final}) of a ground
electronic state diatomic molecule in a weak dc/ac field. Our entire
derivation is based upon the first quantum mechanical principles and
well defined approximations.

The theory developed in this paper is believed to be of importance
for a variety of specific applications, like
e.g.~alignment/orientation of molecules by lasers, trapping of
ultracold molecules in optical lattices, molecular optics and
interferometry, rovibrational spectroscopy of molecules in the
presence of intense laser light, or harmonic generation. Moreover,
the above outlined approach can be extended in a relatively
straightforward manner to the most general case of a polyatomic
molecule interacting with laser light.

\vspace*{+1.00cm}

\end{document}